# Spectroscopy and structural investigation of iron phosphorus trisulfide – FePS$_3$

*Adam K. Budniak\*†, Szymon J. Zelewski\*†, Magdalena Birowska, Tomasz Woźniak, Tatyana Bendikov, Yaron Kauffmann, Yaron Amouyal, Robert Kudrawiec, Efrat Lifshitz\**

Dr. A. K. Budniak, Prof. E. Lifshitz
Schulich Faculty of Chemistry, Solid State Institute, Russell Berrie Nanotechnology Institute, Technion – Israel Institute of Technology, Haifa 3200003, Israel
Emails: abudniak@nus.edu.sg, efrat.l@technion.ac.il

Dr. S. J. Zelewski, Dr. T. Woźniak, Prof. R. Kudrawiec
Department of Semiconductor Materials Engineering, Faculty of Fundamental Problems of Technology, Wrocław University of Science and Technology, Wybrzeże Wyspiańskiego 27, 50-370 Wrocław, Poland
Email: szymon.zelewski@pwr.edu.pl

Dr. M. Birowska
Institute of Theoretical Physics, Faculty of Physics, University of Warsaw, Pasteura 5, 02-093 Warsaw, Poland

Dr. T. Bendikov
Department of Chemical Research Support, Weizmann Institute of Science, 7610001 Rehovot, Israel

Dr. Y. Kauffmann, Prof. Y. Amouyal
Department of Materials Science and Engineering, Technion – Israel Institute of Technology, Haifa 3200003, Israel

†These authors contributed equally: Adam K. Budniak, Szymon J. Zelewski

**Abstract**

Lamellar structures of transition metal phosphorus trisulfides possess strong intralayer bonding, albeit adjacent layers are held by weak van der Waals interactions. Those compounds received enormous interest due to their unique combination of optical and long-range magnetic properties. Among them, iron phosphorus trisulfide (FePS$_3$) gathered special attention for being a semiconductor with an absorption edge in the near-infrared, as well as showing an Ising-like anti-ferromagnetism. We report a successful growth of centimeter size bulk FePS$_3$ crystals with a chemical yield above 70%, whose crystallographic structure and composition were carefully identified by advanced electron microscopy methodologies, including atomic resolution elemental mapping, along with photoelectron spectroscopy. The knowledge on the optical activity of FePS$_3$ is extended utilizing temperature-dependent absorption and photoacoustic spectroscopies, while measurements were corroborated with density-functional theory calculations. Temperature-dependent experiments showed a small and monotonic band-edge energy shift down to 115 K and exposed the interconnected importance of electron-phonon coupling. Most of all, the correlation between the optical behavior and the magnetic phase transition is revealed, which shows the practical utilization of temperature-dependent optical absorption to investigate magnetic interactions.

**1. Introduction**

The rediscovery and intense interest in 2D materials began with the exfoliation of a single graphite layer, called graphene, and presenting its extraordinary electronic[1] and mechanical[2] properties. An entire catalog of such materials emerged[3], dividing them by many criteria, for example, their electronic conductivity (including superconducting states[4,5] and excitonic insulators[6,7]), optical properties, or characteristic element groups forming the crystals which can help predict the character of neighboring compounds by chemical trends[8]. A discovery of bandgap change from indirect to direct in a single layer of MoS$_2$[9,10], and then other transition metal dichalcogenides (TMDs)[11], opened up broad interest in potential applications of semiconducting 2D materials in optoelectronics[12,13]. Prototypes of single-layer devices often exhibit performance superior to the ones fabricated in bulk semiconductors[14,15], at the same time offering tightly confined (<1 nm[14,16]) systems for exploration of quantum effects[17,18].

A unique group of 2D materials receiving growing attention is transition metal phosphorus trichalcogenides[19–21]. They are described by the empirical formula MPX$_3$, where M denotes the transition metal, P phosphorus, and X the chalcogenide. With sulfur being the most common chalcogenide, multiple transition metal trisulfides are formed. Their crystal structure is determined by the hexathiohypo diphosphate anion: $P_2S_6^{4-}$ (simplified as $PS_3^{2-}$) undergoing elongation or distortion to sandwich different transition metals[22]. Many of them reveal a peculiar combination of physicochemical properties: 2D layered structure, semiconducting nature with open bandgap[23,24], and magnetic ordering[25,26] relevant for spintronics and magneto-optics. Optical absorption edges covering the spectral range from near-infrared up to ultraviolet[23,24] raised interest in applications of various $MPX_3$ materials as photodetectors and other opto-electronic devices.

Iron phosphorus trisulfide, $FePS_3$, attracts a lot of attention in that term. It has been shown to exhibit extraordinary photoconductive performance[27–29] arising from high carrier mobility and broadband optical absorption above the bandgap (1.2–1.5 eV depending on the experimental technique[23,24,29]) close to the optimal for solar cells determined by the Shockley–Queisser limit. Negative photoconductivity[28] has also been observed in thin layers of $FePS_3$ for illumination with photons of energy around trap states within the conduction band, switchable in sign and magnitude by applying a gate voltage[27], giving another parameter to control the device performance. Despite that, reports on optical properties of $FePS_3$ at low temperatures, providing details on the typical temperature range of artificially (cryogenics, thermoelectrics) or naturally (space environment[30]) cooled photodetectors, are lacking. Numerous reports cover the magnetic phase transitions in $MPX_3$ compounds based on magnetization measurements[31–33], though the impact on their (including $FePS_3$) optical properties often remains unknown, as the spectroscopic studies are limited to room temperature. Its photosensitivity, competitive among other 2D materials, makes detailed optical and structural studies a necessary step in the development of $MPX_3$-based devices and sensors.

Our work aims to extend the knowledge on the optical activity of $FePS_3$ under conditions typical for cooled, low-noise photodetectors. Bulk $FePS_3$ crystals synthesized with high yield via chemical vapor transport (CVT) were characterized by powder X-ray diffraction (PXRD), scanning electron microscopy (SEM), X-ray photoelectron spectroscopy (XPS), and ultraviolet photoelectron spectroscopy (UPS). Then the crystals were thinned by mechanical exfoliation to confirm the crystallographic structure, employing both transmission electron microscopy (TEM) to determine crystallographic planes, and the high-resolution scanning transmission electron microscopy (HR-STEM) with energy dispersive spectroscopy (EDS) mapping with sub-nanometer resolution for determining atomic arrangement. Then, we performed photoacoustic spectroscopy (PAS) and temperature-dependent optical absorption experiments, revealing a discontinuity of trend in the fundamental absorption edge energy upon the paramagnet to antiferromagnet transition. A *d-d* intra-atomic transition forms another below-gap absorption band, and the analysis of its broadening served as a benchmark parameter for evaluating the electron-phonon coupling strength. The final remarks are complemented by comparing the STEM and optical spectroscopic results with the crystal and band structure calculated in the framework of the density-functional theory (DFT).

## 2. Results and Discussion

### 2.1. Preparation and Characterization of Bulk $FePS_3$

$FePS_3$ single bulk crystals have been synthesized from pure elements in quartz ampoules via chemical vapor transport (CVT)[22,34,35]. The reaction yield exceeded 70% and formed single crystals which approached lateral dimensions of a centimeter (**Figure 1a**). X-ray diffractograms of both ground crystals and a single crystal were recorded (Figure 1b), showing agreement with the diffractions reported previously[36]. **Figure S1** in Supplementary Information (**SI**) introduces the EDS spectrum received with a scanning electron microscope (SEM), confirming the presence of all three elements in the obtained material. A qualitative analysis delivers the atomic ratio Fe : P : S close to 1 : 1 : 3, thus confirming the composition of $FePS_3$. Furthermore, a high-resolution scanning electron microscopy (HR-SEM) image in combination with EDS mapping confirmed the lamellar structure and equal distribution of all elements across the whole crystal (Figures 1c,d).

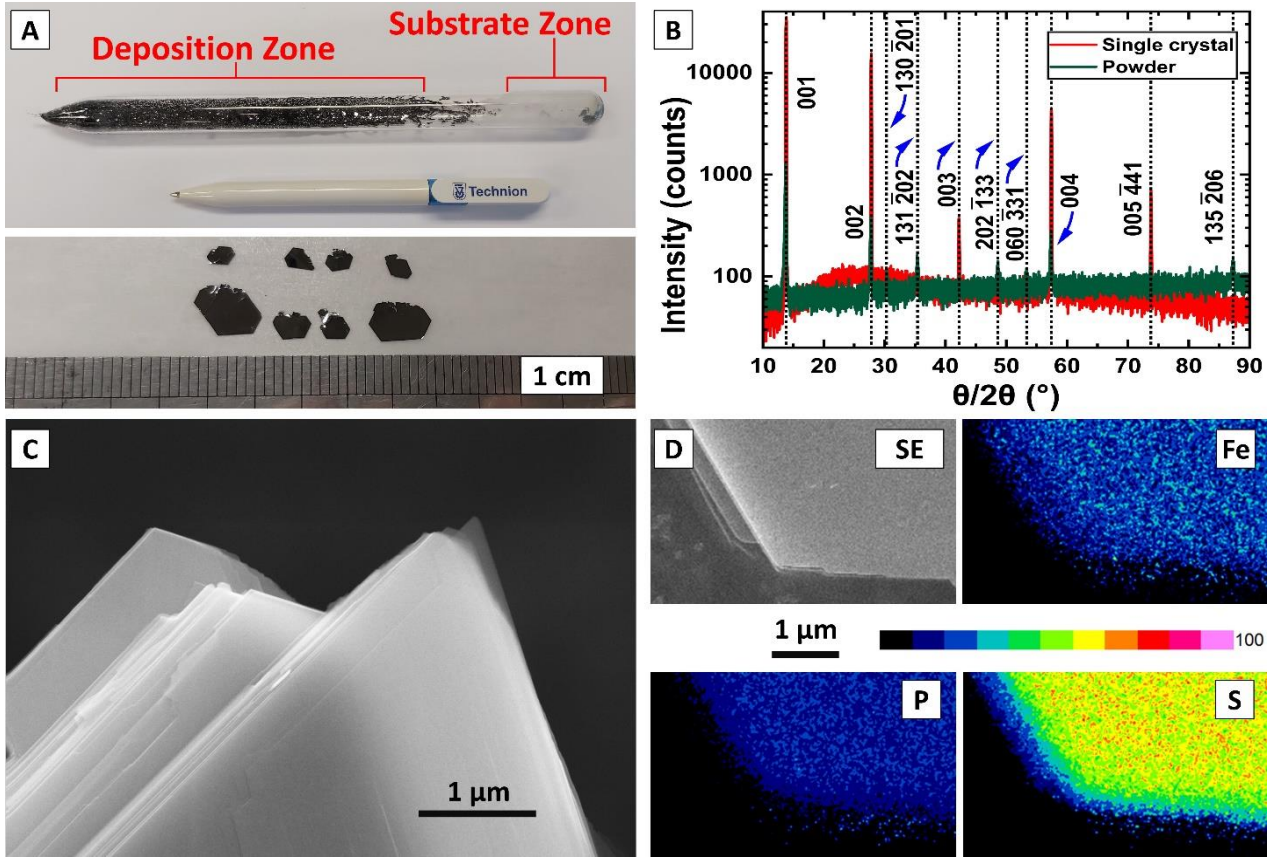

**Figure 1.** A: Photograph of FePS$_3$ flakes as prepared in a sealed quartz ampoule (top) and a few individual flakes out of an ampoule (bottom). B: X-ray diffraction of a powder (green) and a single FePS$_3$ crystal (red). Vertical lines present the position of FePS$_3$ diffraction peaks according to PDF#04-005-1516[36]. C: HR-SEM image of FePS$_3$ single crystal registered by the in-lens detector, exposing the lamellar structure. D: HR-SEM picture registered with secondary electron (SE) detector and three corresponding EDS maps (Fe, P, and S) presenting uniform distribution of the elements across the crystal.

## 2.2. Transmission Electron Microscopy Study of Mechanically Exfoliated FePS$_3$

**Figure S2** in SI presents the FePS$_3$ unit cell along three main directions. FePS$_3$ has monoclinic symmetry *C2/m* with lattice parameters a ≈ 5.947 Å, b ≈ 10.300 Å, c ≈ 6.722 Å, and β ≈ 107.16°[36]. To precisely explore the structural properties of FePS$_3$, a bulk crystal was mechanically exfoliated directly onto a transmission electron microscope grid[37,38]. A detailed description of the mechanical exfoliation and the transfer procedures were given in the SI and also were elaborated in the previous work[37]. Structural models presented in **Figure S3** show geometrical aspects of transmission electron microscopy (TEM) of FePS$_3$ in different zone axes (Z.A.): [001] and [103]. **Figure S4** presents a low magnification micrograph of the whole flake, with the area marked where high-resolution transmission electron microscopy (HR-TEM) and selected area electron diffraction (SAED) experiments were performed. **Figures 2a** and 2c present HR-TEM images after average background subtraction inspected along the [001] and [103] Z.A., respectively. Figures 2b and 2d portray the corresponding SAED patterns, confirming examination at the [001] and [103] Z.A., respectively. For [001] Z.A. electrons are moving along the c direction and three different planes can be described: (200), d = 0.28 nm; (060), d = 0.17 nm and (260), d = 0.15 nm. Although the DP of $FePS_3$ in [001] Z.A. has locally a hexagonal arrangement of diffraction spots, the DP has globally only two-fold symmetry. Consequently, in the HRTEM micrograph of $FePS_3$ in [001] Z.A. only two-fold symmetry is observed. This is in agreement with the crystal structure of $FePS_3$, which is a monoclinic system. Even though $P_2$ pairs and Fe atoms are hexagonally arranged, the $FePS_3$ crystal has two-fold symmetry as sulfur planes lower the symmetry. For [103] Z.A. electrons go along the c* axis, viz., normal to the ab-plane (normal to the surface of the flake). Registration of three equal planes: (060), $(3\bar{3}1)$, and $(33\bar{1})$, with d = 0.17 nm, and a six-fold symmetry (for both HRTEM image and DP) for a monoclinic system, agrees with an observation described by Murayama et al.[39], who explained it by forming of rotational twin structure.

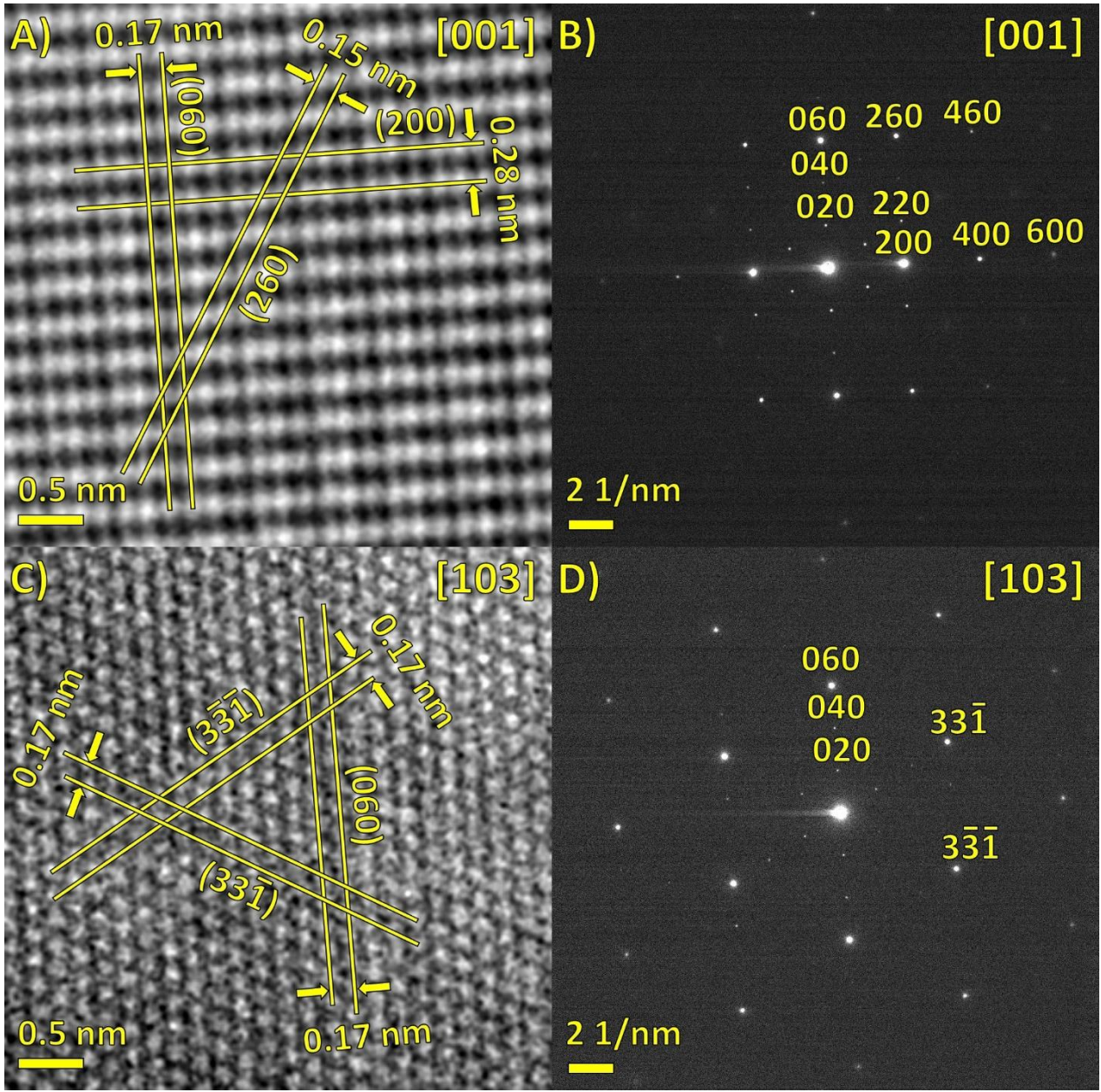

**Figure 2.** TEM experiments for mechanically exfoliated FePS$_3$ crystal, up: at [001] Z.A. and bottom: [103] Z.A. Left (A and C): High-resolution (HR), average background subtraction filtered (ABSF) micrographs with resolved planes. Right (B and D): The corresponding selected area electron diffraction (SAED) patterns. The low magnification micrograph of the FePS$_3$ flake, where all experiments were performed is presented in Figure S4.

Previously, HRTEM experiments were discussed, that usually do not allow observing columns of atoms directly, but lattice fringes. High angle annular dark-field (HAADF) scanning transmission electron microscopy (STEM) has a different image forming mechanism and, for most cases, columns of atoms can be more directly imaged and interpreted.

**Figure 3a** presents a HAADF STEM micrograph of a mechanically exfoliated FePS$_3$ crystal[37,38] along the [001] Z.A. The micrograph is overlapped with a crystallographic model (Fe – red spheres, P – blue spheres, and S – yellow spheres)[36]. Based on the HAADF signal, the mutual distance of 0.35 nm between adjacent Fe–Fe and Fe–P$_2$ columns were determined. Furthermore, Figures 3b–d depict, for the first time, atomically resolved EDS maps of any MPS$_3$ compound acquired along the [001] Z.A. Those maps unprecedently exposed the hexagonal arrangement of the Fe atoms and the columns of phosphorus pairs which are placed in the middle of the Fe hexagons. This conclusion is based on the fact that in the middle of the Fe hexagon, the EDS signal of phosphorus is much higher than that of iron. This observation is in agreement with the crystal structure in this Z.A. The mixing of Fe and P maps can be explained by sample drift, which cannot be avoided at such high magnification. The maps reveal a distance of 0.35 nm between neighboring Fe columns, in agreement with the Fe–Fe distance obtained with the HAADF detector. Moreover, the maps expose a spacing of 0.59 nm between adjacent P$_2$ columns. The described structure and measured inter-atomic distances agree with DFT calculations described further.

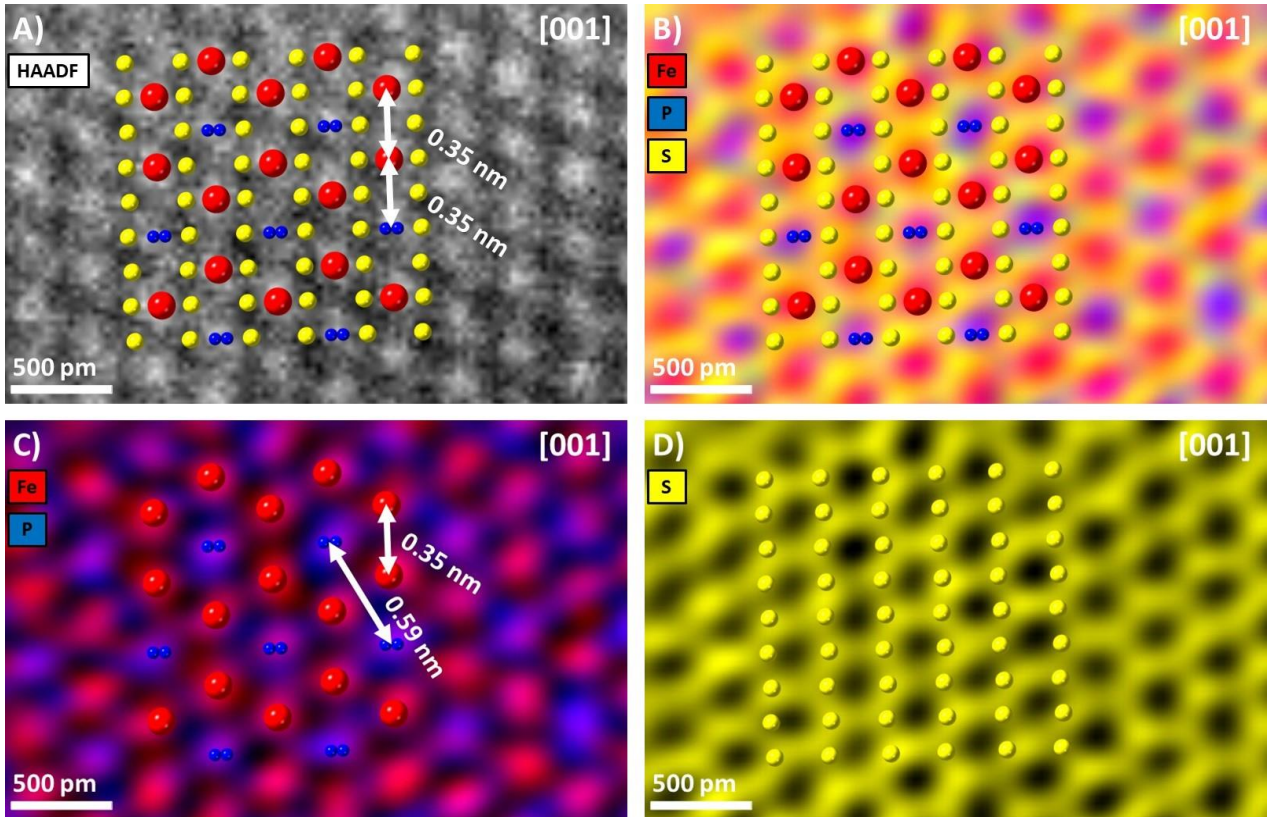

**Figure 3.** HR-STEM experiments for mechanically exfoliated FePS$_3$ at [001] Z.A. A: frames averaged and aligned HR-STEM frames and B–D: filtered atomic resolution EDS elemental maps of FePS$_3$. B: The map presenting all elements (Fe, P, and S). C: presents only Fe and P; where sulfur (S) atoms were omitted for clarity. D: Sulfur (S) EDS elemental maps. All pictures were overlapped with the simulated structure showing the position of the atom columns: Fe (red spheres), P (blue spheres), and S (yellow spheres)[36]. Distances are calculated according to the signal from a detector: HAADF or EDS.

## 2.3. Optical Investigation of Bulk FePS$_3$

The excellent crystallographic quality confirmed by several electron microscopy and X-ray techniques served as a base for choosing a good quality material for optical studies of FePS$_3$. Section 3 and **Figure S5** of the SI present X-ray photoelectron (XPS) and ultraviolet photoemission spectroscopy (UPS) investigations of bulk FePS$_3$. With both techniques (different energy sources, penetration depth, cross-sections for the electronic transitions, etc.), the same values (0.6 eV) of the valence band maximum (VBM) and the second electronic transition (at 2.2 eV, with respect to the Fermi level in Au) were obtained, further strengthening the reliability of these measurements.

**Figure 4a** shows a comparison of room temperature optical absorption and photoacoustic spectra of a bulk FePS$_3$ crystal. Both spectroscopic methods reveal a broad, symmetric absorption feature centered around 1 eV, followed by a steep absorption edge in the 1.4–1.6 eV range. It is worth noting that both methods use monochromatic light illumination with varying wavelengths as opposed to broad-spectrum probing which could cause additional unwanted effects, including excessive sample heating or photovoltaic carrier generation. The photoacoustic spectrum follows the optical absorption profile up to 270 cm$^{-1}$ at 1.4 eV when it saturates, which is a typical effect in photoacoustic detection performed at low modulation frequencies[40]. The phase of the photoacoustic signal undergoes a steep decrease in the vicinity of the amplitude saturation region, related to the heat source appearing closer to the sample surface due to the strong absorption of photons above the bandgap. Photoacoustic spectroscopy does not rely on analysis of light reflected or passed through a studied material and is intrinsically linked with optical absorption, further confirming the nature of any electromagnetic wave extinction below the saturation. The 1 eV absorption peak agrees with the *3d-3d* atomic-like metal ($Fe^{2+}$) transition from fundamental $^5T_{2g}$ to the first excited state $^5E_g$ (further called the *d-d* transition for simplicity), appearing within the forbidden energy region, previously reported on $FePS_3$ and other $MPX_3$ compounds[23,41–43]. Despite the prediction of *d-d* transitions at higher energies[41], derivatives of optical spectra shown in Figure 4b contain only symmetric extrema (each well fitted with a single Gaussian peak) indicating the absence of other hidden transitions, even in the low-temperature spectrum presented in Figure 4c. Oscillations observed in the 0.6 – 1.2 eV region are a result of numerical differentiation and have no physical meaning. The high-energy absorption feature (onset around 1.9 eV) cannot be precisely analyzed (including the temperature dependence) due to significant overlap with the main absorption edge and saturation on the high-energy region resulting from low sample optical transparency.

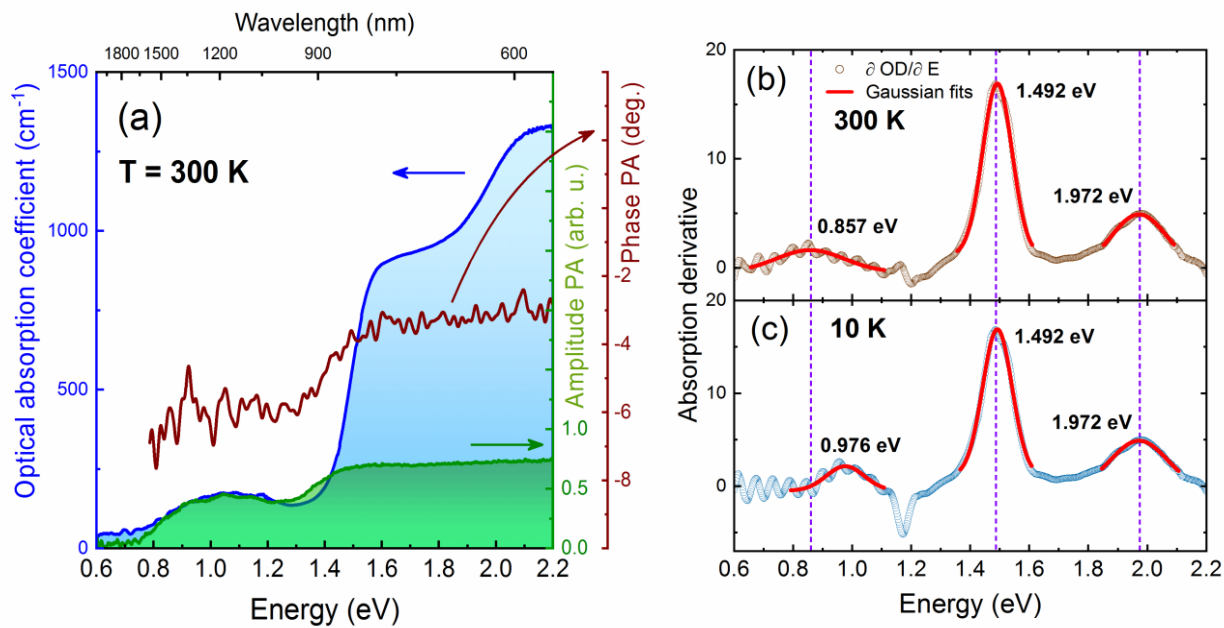

**Figure 4.** (a) Comparison of room temperature optical absorption and photoacoustic spectra deconvoluted to amplitude and phase components. Derivatives of optical absorption spectra revealing edges contributing to light extinction at 300 K (b) and 10 K (c).

The low-energy *d-d* transition can be approximated with a Gaussian shape. As shown in **Figure 5a**, subtraction of the fitted curve from the original spectrum leaves a clean sub-gap absorption residue (Figure 5b). A comparison of Tauc plots for the absorption edge around 1.4 eV proves that the *d-d* transition does not affect the bandgap determination even at room temperature when the transition is the broadest (difference of a few meV), see insets of Figure 5a,b. The temperature dependence of the *d-d* feature shown in Figure 5c reveals a small redshift of the transition peak and a significant decrease of its broadening upon increasing the temperature. The transition energy and full width at half maximum (FWHM) are determined in the full temperature range in Figure 5d, following the procedure previously shown in Figure 5a. The bandgap-related absorption edge shown in Figure 5d also redshifts with the increasing temperature, as revealed by applying Tauc plots to analyze the spectra.

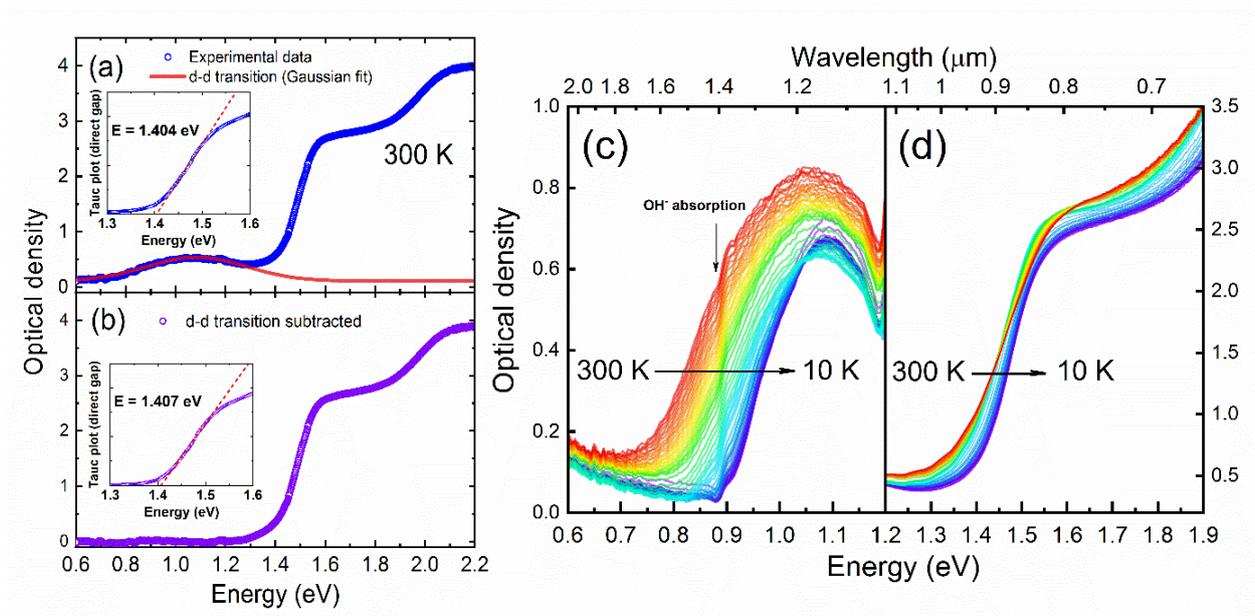

**Figure 5.** (a) Room-temperature optical absorption spectrum with the *d-d* transition approximated with a Gaussian shape, (b) absorption spectrum with the *d-d* transition subtracted; the insets show the Tauc plots used for bandgap determination. (c) Temperature dependence of the *d-d* transition in the infrared range and (d) fundamental absorption edge.

The determined transition energies are summarized in **Figure 6a**. The fundamental absorption edge at 10 K is 1.44 eV and quickly decreases with an abrupt change in the 100–120 K region. The trend is followed by an almost linear shift up to room temperature (~1.40 eV at 300 K), resembling typical models for the temperature dependence of a semiconductor bandgap[44,45]. For comparison, the original Brec's work[23] reports the bandgap energy of $FePS_3$ to be 1.5 eV, though it was determined arbitrarily in the middle of the absorption edge. Our room temperature result is in good agreement with the absorption onset at ~880 nm (~1.41 eV) measured on a thin (11 nm) exfoliated film reported by Gao & Lei et al.[27] Monotonic and predictable absorption edge shift within the temperature range covered by present state-of-the-art thermoelectric cells (200–250 K)[46] is relevant for extending the usefulness of $FePS_3$ as the active material for low-noise cooled photodetectors covering a broad spectral range from near-infrared up to ultraviolet[27–29]. The characteristic region of anomalous temperature dependence coincides with a transition from the paramagnetic (PM) to the antiferromagnetic (AFM) phase of $FePS_3$ (Néel temperature)[47–50]. The same behavior has been observed on a few samples of the synthesized material and during multiple heating/cooling cycles on the same sample, proving good reversibility of the phase transition. Monotonic absorption edge blueshift upon cooling in the range covering the PM-AFM transition has already been observed for other $MPX_3$ materials, including $MnPS_3$[51,52] and $NiPS_3$[53]. This observation might be a link between the magnetic ordering and changes in the band structure of the material resulting from crystal lattice distortion, as previously reported based on temperature-dependent structural studies[39].

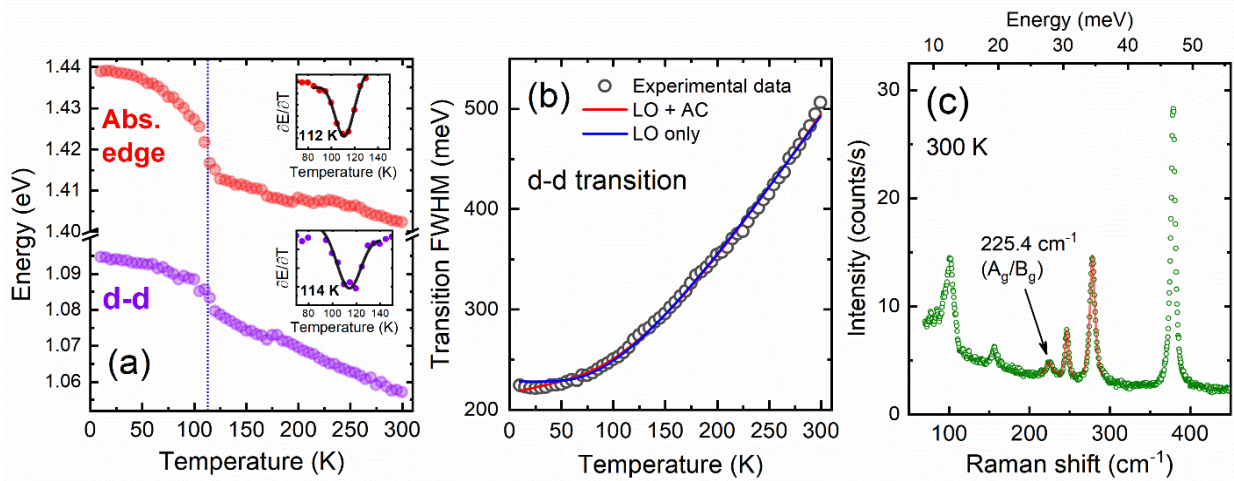

**Figure 6.** (a) Temperature evolution of transition energies for the fundamental absorption edge (red curve) and the *d-d* transition (purple curve). The insets show curve derivatives, with the inflection points attributed to the para- to antiferromagnetic phase transition temperature determined from a single Gaussian fitting. (b) Temperature dependence of the *d-d* transition broadening as determined from the Gaussian approximation with the best approximation evaluating the electron-phonon coupling strength with and without the acoustic phonon contribution. (c) Room temperature Raman scattering spectrum.

The *d-d* transition peak energy, depicted by a purple curve in Figure 6a, goes parallel with the temperature dependence of the fundamental absorption edge, ranging from 1.095 to 1.055 eV. The sudden slope shift around 110 K proves that the localized *d* levels respond to the phase transition via the crystal field effect. Derivatives of both curves and their single Gaussian fits, shown in insets of Figure 6a, reveal more accurately the transition temperatures as (112±1) K for the bandgap absorption and (114±1) K for the *d-d* transition, accordingly. In addition, these results show that the phase transition in $MPX_3$ and other (anti)ferromagnets can be indirectly (through transition-induced changes in crystal parameters) detected by purely optical methods such as simple absorption measurements.

**Figure 6b** shows the temperature dependence of *d-d* transition broadening represented by the FWHM of the fitted Gaussian curve. The huge change from ~220 up to ~500 meV at room temperature suggests strong electron-phonon coupling[54], with various possible carrier scattering mechanisms. By applying the formula:

$$\Gamma(T) = \Gamma_0 + \frac{\gamma_{LO}}{\left(e^{\frac{E_{LO}}{k_B T}} - 1\right)} + \gamma_{AC} T \qquad (1)$$

the contributions of longitudinal optical (LO, Fröhlich) and acoustic (AC) phonons can be distinguished in the broadening parameter. $\Gamma_0$ stands for inhomogeneous broadening, $E_{LO}$ the LO phonon energy, $k_B$ the Boltzmann constant, $\gamma_{LO}$ and $\gamma_{AC}$ the coupling constants for LO and AC phonons, respectively. The FWHM experimental points are fitted with Eq. 1 with and without the acoustic phonon contribution. With the AC coupling parameter left as a free parameter, the fit quality is slightly worse ($R^2$ 0.9981 vs. 0.9997) and does not approximate the data well in the low-temperature range. A low $\gamma_{AC}$ value of $2.4 \times 10^{-4}$ causes the fitted curves to overlap above 100 K. The proper analysis of acoustic phonon contribution in the broadening requires narrow features in optical spectra, typically found in ultra-pure inorganic semiconductors such as GaN[55]. For determining the dominating coupling parameters with LO phonon the $\gamma_{AC}$ is thus omitted. The obtained $E_{LO}$ is (28.1±0.1) meV, in excellent agreement with the 225 cm$^{-1}$ (27.9 meV) $A_g/B_g$ mode found in the Raman spectrum (Figure 6c) well above the phononic bandgap separating the acoustic and optical branches, which also agrees with previously reported results[56,57]. All high-energy peaks (>150 cm$^{-1}$) originate from vibrational modes of the $P_2S_6^{4-}$ anion[58,59], suggesting that electron-phonon coupling strength might be evaluated based on differences in atomic weights of the transition metals[57]. It is relevant to mention that Raman modes above the AFM transition temperature are only affected by intra- and interlayer coupling, from commonly studied transition metal dichalcogenides known to cause energy shifts of particular peaks. In the case of FePS$_3$, mode frequencies between bulk material and thin films down to the monolayer regime do not change significantly, most likely due to relatively weak van der Waals interaction[24,59]. Scarce information about the Fröhlich coupling can be found for MPX$_3$ materials, though recent analysis of temperature-dependent absorption and modulated thermoreflectance spectra have been reported on NiPS$_3$[53]. Significantly higher electron-phonon interaction strength of (1800±800) meV for one of the optical transitions has been explained by flat spin-orbit split-off band dispersion. Our $\gamma_{LO}$ value of (520±2) meV can be explained in a similar matter taking into account the localized character of the *d-d* transition[60] and therefore complements the physical understanding of physical processes affecting the optical activity of transition metal phosphorus trisulfides.

## 2.4. FePS$_3$ Band Structure Calculations

To extend the analysis of optical properties of FePS$_3$ and point out aspects differentiating it from other layered semiconductors, we performed electronic band structure calculations employing the DFT+U method. The FePS$_3$ is an Ising-like antiferromagnet (AFM) with a honeycomb lattice[50]. The magnetic ground state of a bulk system exhibits the AFM zigzag order within the plane, and the adjacent layers are antiferromagnetically aligned[22]. However, the magnetic ground state is not commensurate with the lattice structure, thus, we have chosen the smallest possible magnetic unit cell (see **Figure 7a**) which contains 40 atoms. It is worth mentioning that the position of the band edges depends on the choice of the Hubbard U of the 3d states, which has been shown to have a minor impact in the case of MnPS$_3$[61], but remains significant for FePS$_3$. This difference stems from the fact that in the case of MnPS$_3$ there is a smaller contribution of the 3d states to the band edges than for FePS$_3$. In particular, the Hubbard U affects the position of the Fe 3d states with respect to the Fermi level, which might result in a different bandgap character. Notably, for the U = 2.6 eV, the indirect bandgap is obtained, whereas for the U = 5.3 eV the direct character is clearly visible for the monolayer FePS$_3$ (see **Figure S4**). Thus, the proper Hubbard U is of significant importance (details in SI) and was set to be equal to 2.6 eV, which nicely reproduces the experimental bandgap as well as the positions of the 3d orbital projections similar to the hybrid functional. Our theoretical results for the FePS$_3$ bulk structure reveal an indirect character of the electronic bandgap, which is equal to 1.44 eV (see **Figure 7**) and matches perfectly the experimental bandgap of 1.4 eV obtained here from several optical measurements. Nevertheless, the bands close to the Fermi level are flat and the difference between the indirect and direct transitions is 60 meV. In addition, a much smaller bandgap equal to 1.23 eV has been previously reported for the bulk system[29], which is consistent with the choice of the U (U = 2.2 eV). However, the transition reported in this work occurs at different k-points in the Brillouin zone (BZ), which points to a different unit cell, not commensurate with the magnetic unit cell of a bulk (see **Figure 7a**). In addition, the comparison between the monolayer (ML) and bulk system reveals that the theoretical bandgap of bulk is 0.4 eV smaller than for the ML case (U = 2.6 eV, please refer to SI). This is in line with the general trend of the bandgap behavior for the van der Waals materials.

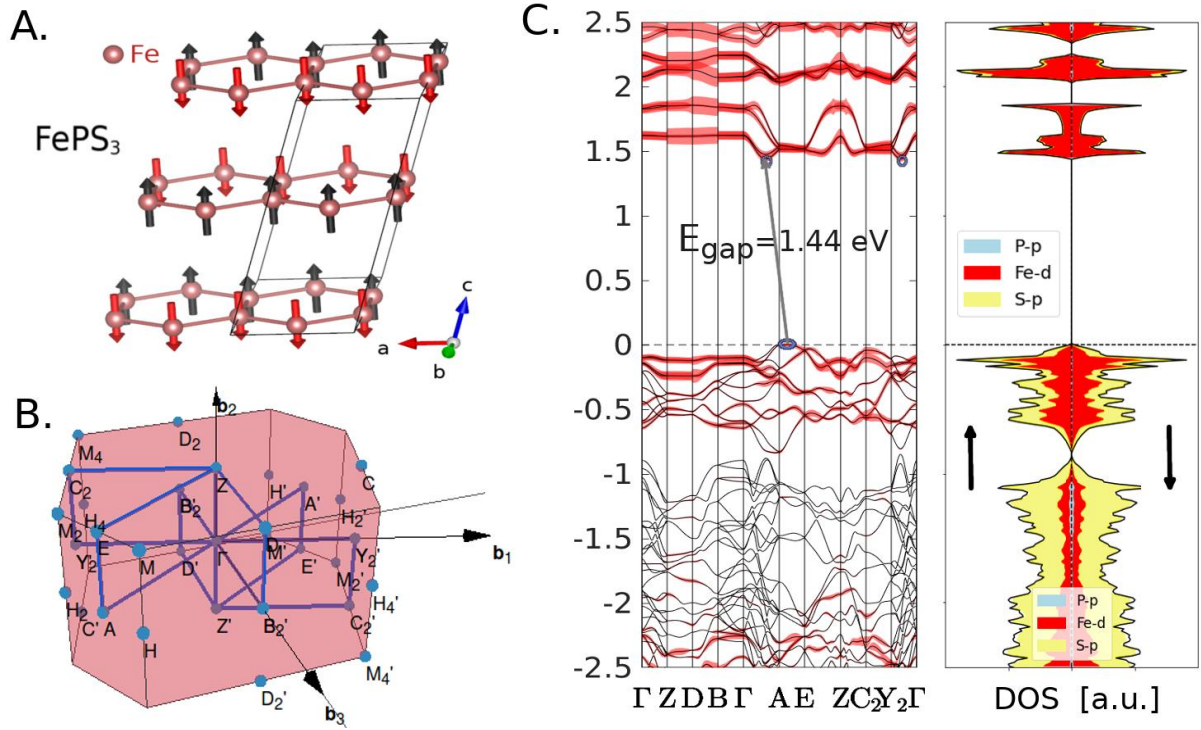

**Figure 7.** Electronic structure of the bulk FePS$_3$ obtained with GGA + D3 + U (U = 2.6 eV) approach. (A) A magnetic unit cell (solid black lines) used in the calculations (for clarity only iron atoms are shown) with a (B) corresponding first BZ with high symmetry k-points and recommended high symmetry k-path[62] are plotted in blue. (C) The fundamental indirect bandgap equal to 1.44 eV is visible. The difference between the direct and indirect transitions is 60 meV and may depend on the Hubbard U. The 3d band projections of the Fe atoms denoted in the red and projected density of states (PDOS) are presented. The main contribution close to the Fermi level comes from the 3d states of Fe atoms. The valence and conduction band edges are denoted by blue circles. Note that the two conduction band minima are visible (two valleys). The arrows in (A) indicate the spin arrangements on iron atoms.

The structural parameters, as well as distances for the bulk system, are in good agreement with our structural characterization (compared in **Figure 8**) and with previously reported *ab initio* results[63].

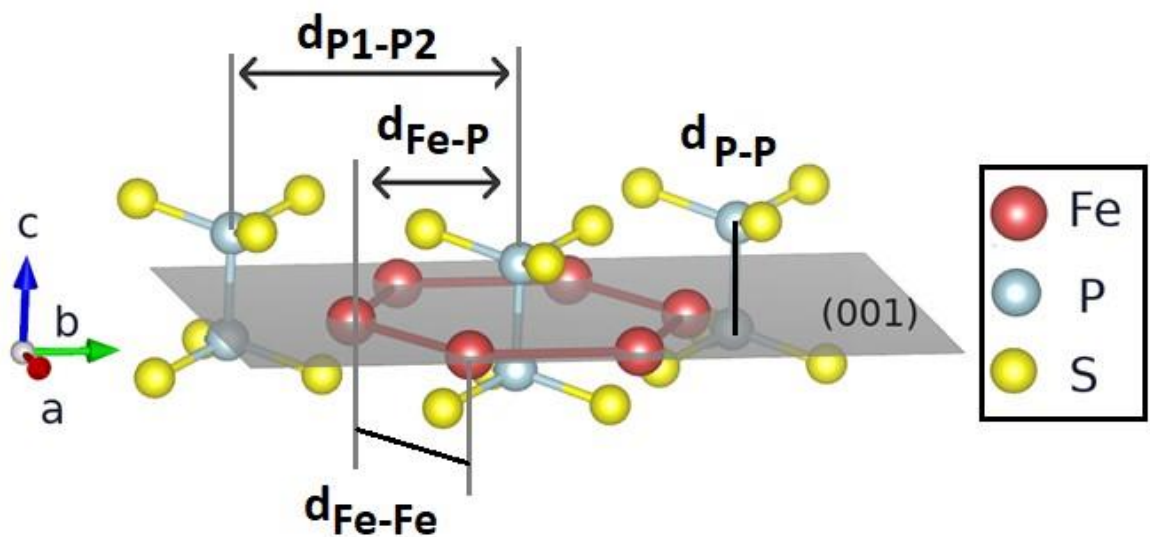

| | HR-STEM experiment (Å) | DFT calculations (Å) |
|---|---|---|
| $d_{Fe-Fe}$ | 3.5 | 3.438 |
| $d_{P-P}$ | - | 2.204 |
| $d_{P1-P2}$ | 5.9 | 5.969 |
| $d_{Fe-P}$ | 3.5 | 3.591 |

**Figure 8.** Side view model of FePS$_3$ monolayer: Fe (red spheres), P (blue spheres), and S (yellow spheres). The (001) plane is presented also with inter-atom distances that are discussed in the table. The table introduces a comparison between experimental (atomic resolution EDS mapping and HR-STEM imaging presented in Figure 3) and theoretical distances between the atoms. Structural parameters for the bulk obtained in the GGA + D3 + U approach are in good agreement with previously reported values[63].

## 3. Conclusions

We demonstrated a chemical vapor transport synthesis of large, centimeter size $FePS_3$ crystals, providing 70% product yield. The extensive structural characterization confirms the compositional uniformity of the synthesized material. To the best of our knowledge, for the first time on any transition metal phosphorus trisulfide, different positions of atomic columns were revealed by high-resolution STEM coupled with EDS mapping. $FePS_3$ was further investigated independently via XPS and UPS, showing a valence band maximum for 0.6 eV and a second electronic transition at 2.2 eV. The optical activity of $FePS_3$ has been evaluated at room temperature by photoacoustic and optical spectroscopy, distinguishing the *d-d* intraatomic-like transition from the fundamental absorption edge. Temperature-dependent absorption spectra provide new insight into the potential usefulness of $FePS_3$ as the active material in cooled photodetectors, as the bandgap energy (1.4 eV) shift is monotonic and relatively small down to the magnetic phase transition temperature around 115 K. Additionally, the demonstrated fully optical approach to study phase transitions in (anti)ferromagnetic materials reveals the link between magnetic arrangement and band structure changes, paving the way toward future explorations of magneto-optic coupling effects. Huge absorption broadening sheds new light on the localized nature of the observed *d-d* transition and leads to an analysis of electron-optical phonon interaction strength. The calculated band structure within the DFT + U approach, first used to verify the findings of optical spectroscopy, points out the relevance of considering a proper Hubbard parameter (U) value on electronic structure results, affecting the band edges and the determined bandgap character. We believe that our results will support future research and increase the interest in iron phosphorus trisulfide as a representative of unique magnetic layered semiconductors.

## 4. Experimental Section

All experimental details and methodology are given in the Supplementary Information (SI). VESTA software[64] was used for crystal structure visualization, including the SI.

**Data availability**

All data derived from the experiments and calculations of this study are available from the authors upon reasonable request.

**Supporting Information**

Supporting Information is presented in the further section of this manuscript.

**Author contributions**

A. K. B. synthesized materials and performed structural experiments (XRD, electron microscopy). S. J. Z. did temperature-dependent absorption and photoacoustic spectroscopy experiments. M. B. implemented DFT calculations for monolayer and bulk material with T. W. support. Y. K. with A. K. B. registered atomic resolution EDS-STEM maps. T. B. did XPS and UPS experiments. R. K. supervised optical measurements. E. L. and Y. A. superintended materials preparation. S. J. Z., A. K. B., M. B., and T. B. interpreted results and wrote the manuscript with input from all authors.

**Acknowledgments**

A. K. B. and E. L. were supported by the European Commission via the Marie Skłodowska-Curie action Phonsi (H2020-MSCA-ITN-642656). S. J. Z. and R. K. performed research within the grant of the National Science Centre Poland (OPUS 11 no. 2016/21/B/ST3/00482). S. J. Z. is a beneficiary of the START scholarship from the Foundation for Polish Science. M. B. acknowledges financial support from the National Science Centre Poland (SONATA 12 grant no. 2016/23/D/ST3/03446). Access to computing facilities of TU Dresden (ZIH) within the project "TransPheMat", PL-Grid Polish Infrastructure for Supporting Computational Science in the European Research Space, and of the Interdisciplinary Center of Modeling (ICM), University of Warsaw are gratefully acknowledged. E. L. acknowledges the financial support from the Israel Science Foundation (No. 2528/19) and from the USA National Science Foundation–USA/Israel Binational Science Foundation (NSF-BSF, No. 2017637).

We thank Prof. Paweł Scharoch (Wrocław University of Science and Technology, Poland) for fruitful discussions. We want to acknowledge Prof. Reshef Tenne, Dr. Rita Rosentsveig, and Dr. Marco Serra (Weizmann Institute of Science, Israel) with an appreciation for the introduction to the synthesis of $FePS_3$. We thank Dr. Alex Berner (Technion – Israel Institute of Technology, Israel) for advice for SEM microscopy, and Mr. Michael Kalina (Technion – Israel Institute of Technology, Israel) for his technical support.

**Competing interests:** The authors declare no competing interests.

**ORCID IDs**

Adam K. Budniak: 0000-0001-7073-8279

Szymon J. Zelewski: 0000-0002-6037-3701

Magdalena Birowska: 0000-0001-6357-7913

Tomasz Woźniak: 0000-0002-2290-5738

Tatyana Bendikov: 0000-0002-1637-6366

Yaron Kauffmann: 0000-0002-0117-6222

Yaron Amouyal: 0000-0003-2198-3539

Robert Kudrawiec: 0000-0003-2593-9172

Efrat Lifshitz: 0000-0001-7387-7821

**References**

[1] K. S. Novoselov, A. K. Geim, S. V. Morozov, D. Jiang, Y. Zhang, S. V. Dubonos, I. V. Grigorieva, A. A. Firsov, *Science* **2004**, *306*, 666.

[2] C. Lee, X. Wei, J. W. Kysar, J. Hone, *Science* **2008**, *321*, 385.

[3] P. Miró, M. Audiffred, T. Heine, *Chem. Soc. Rev.* **2014**, *43*, 6537.

[4] Y. Saito, T. Nojima, Y. Iwasa, *Nat. Rev. Mater.* **2017**, *2*, 16094.

[5] D. Qiu, C. Gong, S. Wang, M. Zhang, C. Yang, X. Wang, J. Xiong, *Adv. Mater.* **2021**, *33*, 2006124.

[6] Y. Wakisaka, T. Sudayama, K. Takubo, T. Mizokawa, M. Arita, H. Namatame, M. Taniguchi, N. Katayama, M. Nohara, H. Takagi, *Phys. Rev. Lett.* **2009**, *103*, 026402.

[7] L. Li, W. Wang, L. Gan, N. Zhou, X. Zhu, Q. Zhang, H. Li, M. Tian, T. Zhai, *Adv. Funct. Mater.* **2016**, *26*, 8281.

[8] M. A. Susner, M. Chyasnavichyus, M. A. McGuire, P. Ganesh, P. Maksymovych, *Adv. Mater.* **2017**, *29*, 1602852.

[9] A. Splendiani, L. Sun, Y. Zhang, T. Li, J. Kim, C. Y. Chim, G. Galli, F. Wang, *Nano Lett.* **2010**, *10*, 1271.

[10] K. F. Mak, C. Lee, J. Hone, J. Shan, T. F. Heinz, *Phys. Rev. Lett.* **2010**, *105*, 136805.

[11] W. Shi, M.-L. Lin, Q.-H. Tan, X.-F. Qiao, J. Zhang, P.-H. Tan, *2D Mater.* **2016**, *3*, 025016.

[12] W. Zhang, Q. Wang, Y. Chen, Z. Wang, A. T. S. Wee, *2D Mater.* **2016**, *3*, 022001.

[13] K. F. Mak, J. Shan, *Nat. Photonics* **2016**, *10*, 216.

[14] B. Radisavljevic, A. Radenovic, J. Brivio, V. Giacometti, A. Kis, *Nat. Nanotechnol.* **2011**, *6*, 147.

[15] O. Lopez-Sanchez, D. Lembke, M. Kayci, A. Radenovic, A. Kis, *Nat. Nanotechnol.* **2013**, *8*, 497.

[16] S. B. Desai, S. R. Madhvapathy, A. B. Sachid, J. P. Llinas, Q. Wang, G. H. Ahn, G. Pitner, M. J. Kim, J. Bokor, C. Hu, H.-S. P. Wong, A. Javey, *Science* **2016**, *354*, 99.

[17] C. Chakraborty, K. M. Goodfellow, S. Dhara, A. Yoshimura, V. Meunier, A. N. Vamivakas, *Nano Lett.* **2017**, *17*, 2253.

[18] M. Parzefall, Á. Szabó, T. Taniguchi, K. Watanabe, M. Luisier, L. Novotny, *Nat. Commun.* **2019**, *10*, 292.

[19] L. Wang, P. Hu, Y. Long, Z. Liu, X. He, *J. Mater. Chem. A* **2017**, *5*, 22855.

[20] R. Samal, G. Sanyal, B. Chakraborty, C. S. Rout, *J. Mater. Chem. A* **2021**, *9*, 2560.

[21] Y. Zhang, T. Fan, S. Yang, F. Wang, S. Yang, S. Wang, J. Su, M. Zhao, X. Hu, H. Zhang, T. Zhai, *Small Methods* **2021**, *5*, 2001068.

[22] R. Brec, *Solid State Ion.* **1986**, *22*, 3.

[23] R. Brec, D. M. Schleich, G. Ouvrard, A. Louisy, J. Rouxel, *Inorg. Chem.* **1979**, *18*, 1814.

[24] K. Z. Du, X. Z. Wang, Y. Liu, P. Hu, M. I. B. Utama, C. K. Gan, Q. Xiong, C. Kloc, *ACS Nano* **2016**, *10*, 1738.

[25] M. Gibertini, M. Koperski, A. F. Morpurgo, K. S. Novoselov, *Nat. Nanotechnol.* **2019**, *14*, 408.

[26] M. Blei, J. L. Lado, Q. Song, D. Dey, O. Erten, V. Pardo, R. Comin, S. Tongay, A. S. Botana, *Appl. Phys. Rev.* **2021**, *8*, 021301.

[27] Y. Gao, S. Lei, T. Kang, L. Fei, C. L. Mak, J. Yuan, M. Zhang, S. Li, Q. Bao, Z. Zeng, Z. Wang, H. Gu, K. Zhang, *Nanotechnology* **2018**, *29*, 244001.

[28] Z. Ou, T. Wang, J. Tang, X. Zong, W. Wang, Q. Guo, Y. Xu, C. Zhu, L. Wang, W. Huang, H. Xu, *Adv. Opt. Mater.* **2020**, *8*, 2000201.

[29] M. Ramos, F. Carrascoso, R. Frisenda, P. Gant, S. Mañas-Valero, D. L. Esteras, J. J. Baldoví, E. Coronado, A. Castellanos-Gomez, M. R. Calvo, *NPJ 2D Mater. Appl.* **2021**, *5*, 1.

[30] R. Thirsk, A. Kuipers, C. Mukai, D. Williams, *Can. Med. Assoc. J.* **2009**, *180*, 1216.

[31] C. C. Mayorga-Martinez, Z. Sofer, D. Sedmidubský, Š. Huber, A. Y. S. Eng, M. Pumera, *ACS Appl. Mater. Interfaces* **2017**, *9*, 12563.

[32] Y. Peng, S. Ding, M. Cheng, Q. Hu, J. Yang, F. Wang, M. Xue, Z. Liu, Z. Lin, M. Avdeev, Y. Hou, W. Yang, Y. Zheng, J. Yang, *Adv. Mater.* **2020**, *32*, 2001200.

[33] M. J. Coak, D. M. Jarvis, H. Hamidov, A. R. Wildes, J. A. M. Paddison, C. Liu, C. R. S. Haines, N. T. Dang, S. E. Kichanov, B. N. Savenko, S. Lee, M. Kratochvílová, S. Klotz, T. C. Hansen, D. P. Kozlenko, J.-G. Park, S. S. Saxena, *Phys. Rev. X* **2021**, *11*, 011024.

[34] W. Klingen, R. Ott, H. Hahn, *ZAAC - Journal of Inorganic and General Chemistry* **1973**, *396*, 271.

[35] B. E. Taylor, J. Steger, A. Wold, *J. Solid State Chem.* **1973**, *7*, 461.

[36] G. Ouvrard, R. Brec, J. Rouxel, *Mater. Res. Bull* **1985**, *20*, 1181.

[37] A. K. Budniak, N. A. Killilea, S. J. Zelewski, M. Sytnyk, Y. Kauffmann, Y. Amouyal, R. Kudrawiec, W. Heiss, E. Lifshitz, *Small* **2020**, *16*, 1905924.

[38] M. Shentcis, A. K. Budniak, X. Shi, R. Dahan, Y. Kurman, M. Kalina, H. Herzig Sheinfux, M. Blei, M. K. Svendsen, Y. Amouyal, S. Tongay, K. S. Thygesen, F. H. L. Koppens, E. Lifshitz, F. J. García de Abajo, L. J. Wong, I. Kaminer, *Nat. Photonics* **2020**, *14*, 686.

[39] C. Murayama, M. Okabe, D. Urushihara, T. Asaka, K. Fukuda, M. Isobe, K. Yamamoto, Y. Matsushita, *J. Appl. Phys.* **2016**, *120*, 142114.

[40] P. Poulet, J. Chambron, R. Unterreiner, *J. Appl. Phys.* **1980**, *51*, 1738.

[41] M. Piacentini, F. S. Khumalo, G. Leveque, C. G. Olson, D. W. Lynch, *Chem. Phys.* **1982**, *72*, 61.

[42] V. Grasso, S. Santangelo, M. Piacentini, *Solid State Ion.* **1986**, *20*, 9.

[43] E. J. K. B. Banda, *phys. stat. sol. (b)* **1986**, *138*, K125.

[44] Y. P. Varshni, *Physica* **1967**, *34*, 149.

[45] K. P. O'Donnell, X. Chen, *Appl. Phys. Lett.* **1991**, *58*, 2924.

[46] D. Zhao, G. Tan, *Appl. Therm. Eng.* **2014**, *66*, 15.

[47] G. Le Flem, R. Brec, G. Ouvard, A. Louisy, P. Segransan, *J. Phys. Chem. Solids* **1982**, *43*, 455.

[48] K. Kurosawa, S. Saito, Y. Yamaguchi, *J. Phys. Soc. Japan* **1983**, *52*, 3919.

[49] P. A. Joy, S. Vasudevan, *Phys. Rev. B* **1992**, *46*, 5425.

[50] D. Lançon, H. C. Walker, E. Ressouche, B. Ouladdiaf, K. C. Rule, G. J. McIntyre, T. J. Hicks, H. M. Rønnow, A. R. Wildes, *Phys. Rev. B* **2016**, *94*, 214407.

[51] V. Grasso, F. Neri, P. Perillo, L. Silipigni, M. Piacentini, *Phys. Rev. B* **1991**, *44*, 11060.

[52] V. G. Piryatinskaya, I. S. Kachur, V. V. Slavin, A. V. Yeremenko, Yu. M. Vysochanskii, *Low Temp. Phys.* **2012**, *38*, 870.

[53] C.-H. Ho, T.-Y. Hsu, L. C. Muhimmah, *NPJ 2D Mater. Appl.* **2021**, *5*, 8.

[54] A. D. Wright, C. Verdi, R. L. Milot, G. E. Eperon, M. A. Pérez-Osorio, H. J. Snaith, F. Giustino, M. B. Johnston, L. M. Herz, *Nat. Commun.* **2016**, *7*, 11755.

[55] A. K. Viswanath, J. I. Lee, *Phys. Rev. B* **1998**, *58*, 16333.

[56] M. Bernasconi, G. Benedek, L. Miglio, *Phys. Rev. B* **1988**, *38*, 12100.

[57] F. Kargar, E. A. Coleman, S. Ghosh, J. Lee, M. J. Gomez, Y. Liu, A. S. Magana, Z. Barani, A. Mohammadzadeh, B. Debnath, R. B. Wilson, R. K. Lake, A. A. Balandin, *ACS Nano* **2020**, *14*, 2424.

[58] M. Scagliotti, M. Jouanne, M. Balkanski, G. Ouvrard, G. Benedek, *Phys. Rev. B* **1987**, *35*, 7097.

[59] J. U. Lee, S. Lee, J. H. Ryoo, S. Kang, T. Y. Kim, P. Kim, C. H. Park, J. G. Park, H. Cheong, *Nano Lett.* **2016**, *16*, 7433.

[60] R. Atta-Fynn, P. Biswas, D. A. Drabold, *Phys. Rev. B* **2004**, *69*, 245204.

[61] M. Birowska, P. E. Faria Junior, J. Fabian, J. Kunstmann, *Phys. Rev. B* **2021**, *103*, L121108.

[62] Y. Hinuma, G. Pizzi, Y. Kumagai, F. Oba, I. Tanaka, *Comput. Mater. Sci.* **2017**, *128*, 140.

[63] B. L. Chittari, Y. Park, D. Lee, M. Han, A. H. Macdonald, E. Hwang, J. Jung, *Phys. Rev. B* **2016**, *94*, 184428.

[64] K. Momma, F. Izumi, *J. Appl. Crystallogr.* **2011**, *44*, 1272.

**Table of contents text**

Iron phosphorus trisulfide (FePS$_3$) receives great attention as a layered semiconductor with a bandgap in the infrared region, simultaneously exhibiting antiferromagnetism at cryogenic temperatures. Using photoacoustic and absorption experiments, we identify optical transitions and their temperature evolution, revealing the influence of the magnetic phase transition on optical properties of FePS$_3$ and electron-phonon coupling strength, relevant for magnetooptic and light detection applications.

**Table of contents figure**

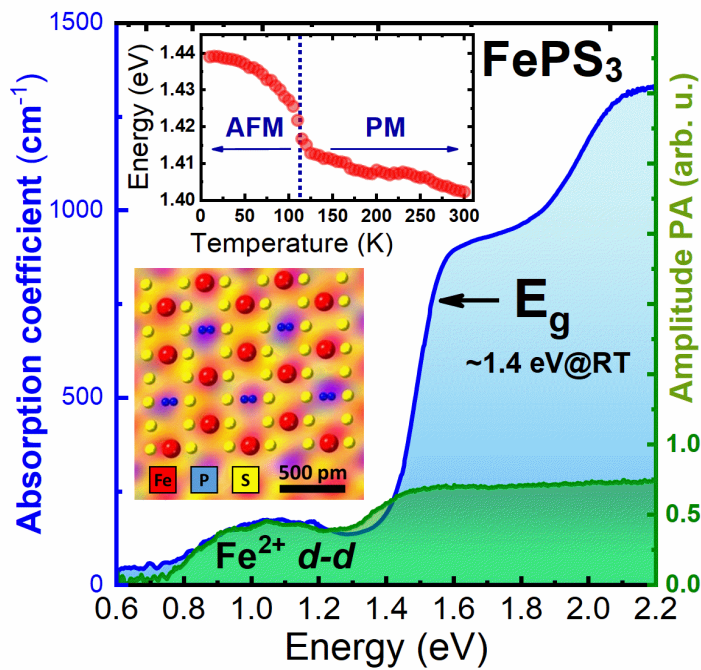

Supporting Information

**Spectroscopy and structural investigation of iron phosphorus trisulfide – FePS$_3$**

*Adam K. Budniak\*†, Szymon J. Zelewski\*†, Magdalena Birowska, Tomasz Woźniak, Tatyana Bendikov, Yaron Kauffmann, Yaron Amouyal, Robert Kudrawiec, Efrat Lifshitz\**

# 1. SEM with EDS

**Figure S1** presents the EDS spectrum of bulk FePS$_3$ as measured with scanning electron microscopy (SEM). The figure designates the existence of all three elements: iron, phosphorus, and sulfur. The inset presents a backscattered electron (BSE) micrograph of a single crystal which exposes a lamellar structure. The table presents the atomic composition, with Fe : P : S ratio close to 1 : 1 : 3, thus confirming the FePS$_3$ stoichiometry.

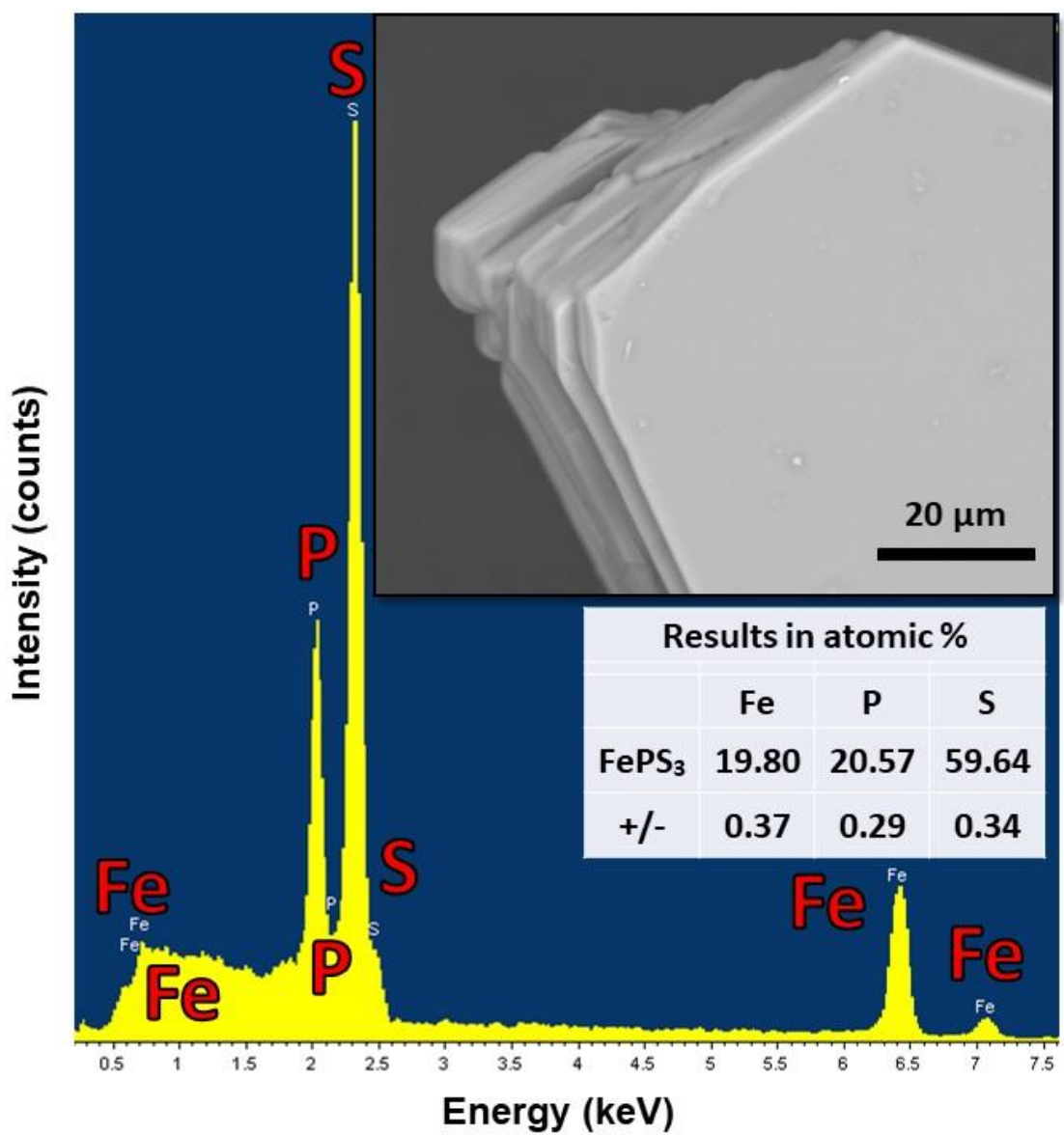

**Figure S1.** An EDS-SEM spectrum of FePS$_3$ single crystal. The top inset displays a backscattered electron micrograph; the bottom inset supplies the atomic composition percentage according to the EDS spectrum.

## 2. FePS$_3$ structural models
**Figure S2** presents FePS$_3$ structural models along main crystallographic directions.

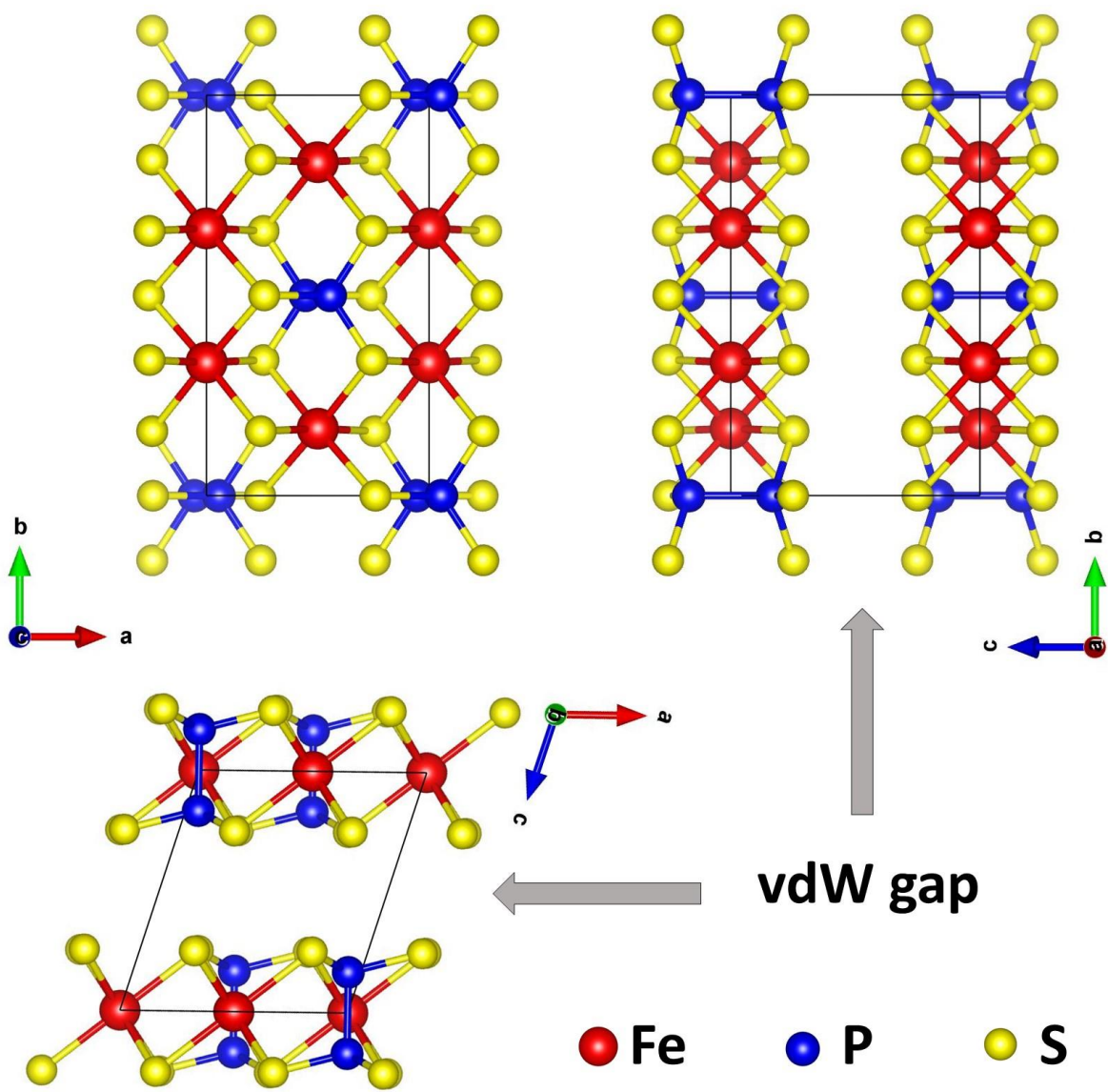

**Figure S2.** FePS$_3$ unit cell for three basic directions[1]. Iron atoms are represented by red spheres, phosphorus atoms by blue spheres, and sulfur atoms by yellow spheres. Solid black lines show the unit cell and grey arrows point to the vdW gap.

## 3. Transmission electron microscopy (TEM)

**Figure S3** below presents the orientation of the FePS$_3$ flake with respect to the electron beam during TEM investigation, when the crystal is aligned in [001] Z.A. and in [103] Z.A., respectively.

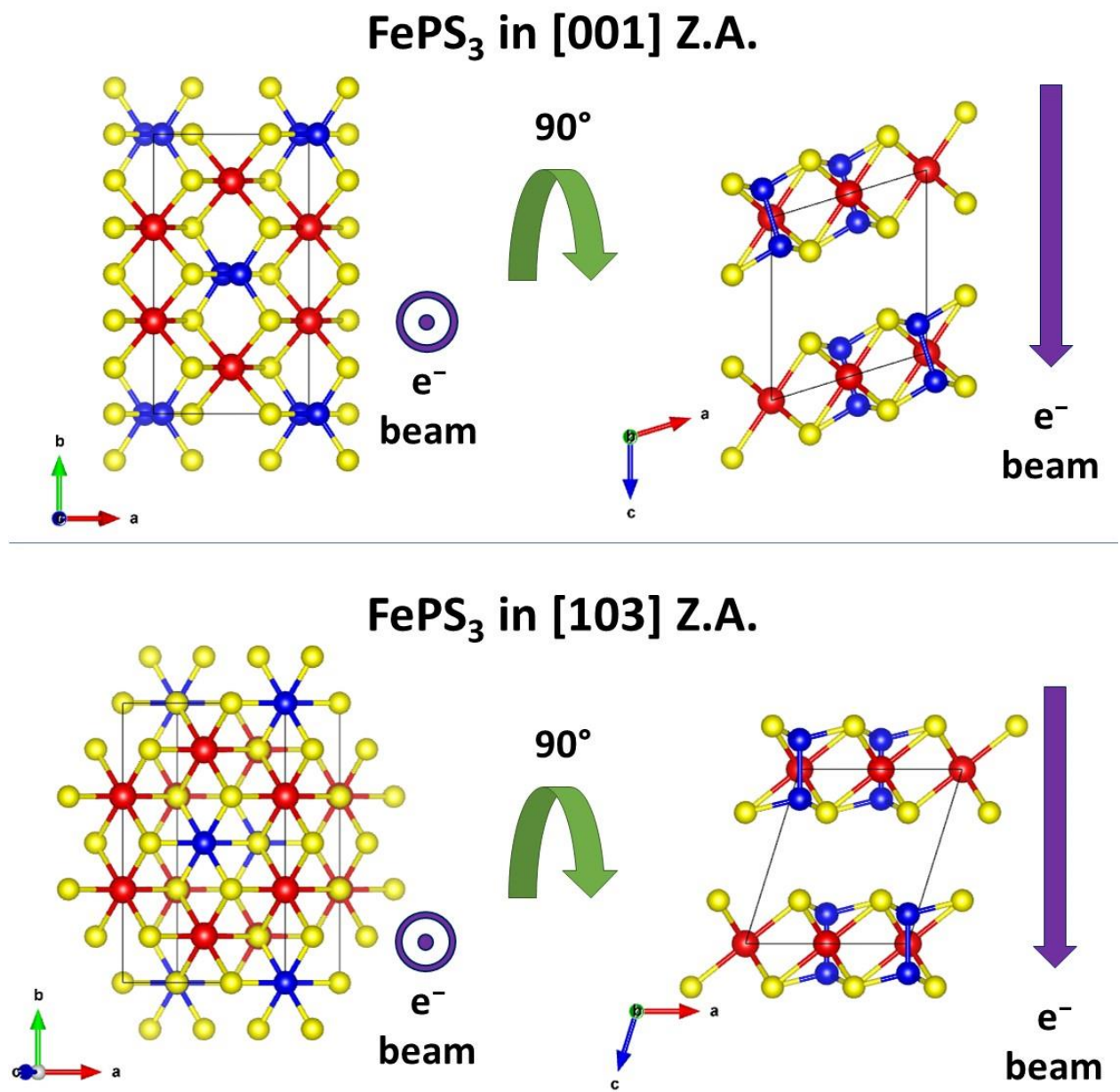

**Figure S3.** Schematic models presenting orientation of the FePS$_3$ crystal during TEM investigation for [001] Z.A. and [103] Z.A.[1] Iron atoms are represented by red spheres, phosphorus atoms by blue spheres, and sulfur atoms by yellow spheres. Solid black lines show the unit cell.

FePS$_3$ crystal was directly exfoliated onto a transmission electron microscopy grid[2,3]. **Figure S4** below presents a low magnification picture of the whole FePS$_3$ flake, with the marked area, where HR-TEM and SAED experiments were performed (**Figure 2**).

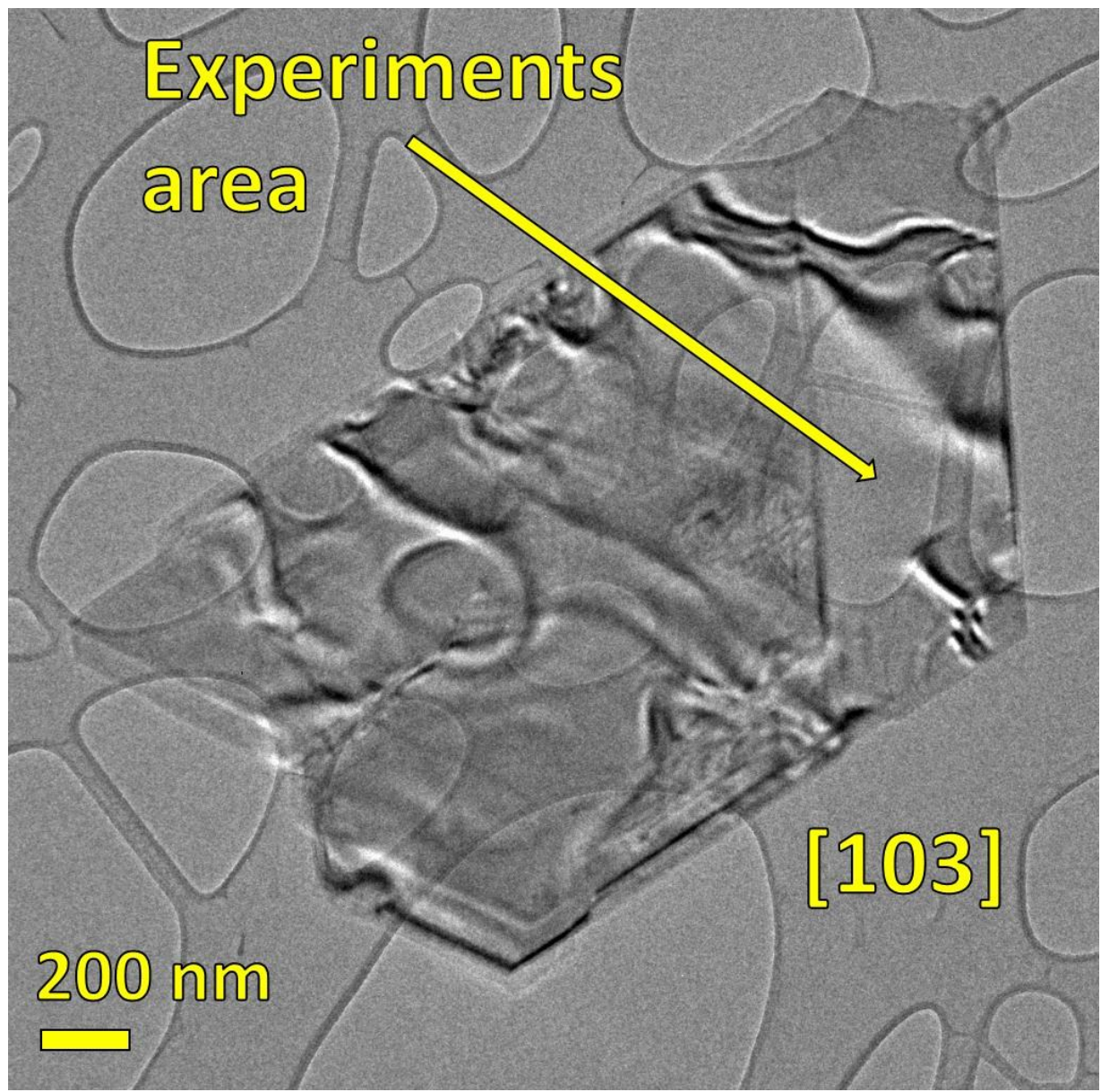

**Figure S4:** TEM micrograph of a mechanically exfoliated FePS$_3$ crystal, registered at the [103] zone axis. The arrow marks the area where observations depicted in Figure 2 were registered.

# 4. X-ray photoelectron (XPS) and ultraviolet photoelectron (UPS) spectroscopies

The chemical composition of $FePS_3$ was examined by X-ray photoelectron spectroscopy (XPS) measurements. For this, ~5×5 mm $FePS_3$ flake (as-synthesized) was attached to the carbon tape strip and loaded into the spectrometer. High-resolution XPS spectra of Fe 2p, P 2p, S 2p, and valence band regions are shown in **Figure S5**: The Fe 2p region (A) includes Fe $2p_{3/2}$ component with a central peak at 708.6 eV, typical of a Fe(II) paramagnetic shape, coinciding with the $FePS_3$ characteristic[4,5]. The P $2p_{3/2}$ and S $2p_{3/2}$ components in panels Figure S5B and S5C, are related to P and S of the $FePS_3$ compound, respectively[4–6]. The last two panels show also higher energy bands (between 133 to 166 eV) that may be related to a few different oxidized components (e.g., $S_xO_y$, $P_xO_y$) or/and to carbon bonded moieties (e.g., S-C, P-C). Those contaminations may have been created by short exposure of the $FePS_3$ flakes to ambient conditions prior to the XPS measurement, a period which allowed adsorption of foreign species, which are strongly pronounced by a surface related methodology such as XPS.

Quantitative analysis of the XPS spectra shows atomic ratios of Fe : P : S = 0.75 : 1 : 2.55 between the elements, which is different from the theoretically expected ratios and those obtained from EDS-SEM (Fe : P : S close to 1 : 1 : 3, Figure S1). The main reason for these differences is the sensitivity of the method to the surface. With EDS-SEM main signal obtained from the depth of 100–200 nm of the sample and thus represents elemental composition in the sample bulk, while in XPS 95% of the signal comes from the top 3 nm, thus representing the elemental composition of the surface, which is not necessarily the same as the composition in the bulk.

The high-resolution spectrum of the valence band region measured with the X-ray source is shown in Figure S5D. The blue line corresponds to a baseline control experiment. The first valence electron transition appeared at 0.6 eV and this value disagrees with previous reports. In 1982 Piacentini et al. reported the first electron transition for $FePS_3$ at 0.8 eV[7]. Later, in 1996, Zhukov et al. theoretically predicted a value of 0.9 eV[8]. A sparse of values may be related to a variation of the Fermi level determination.

Details of the electronic structure of $FePS_3$ were further investigated by ultraviolet photoelectron spectroscopy (UPS) measurements. A work function ($W_f$) value of 4.17 eV was calculated from the high secondary electron photoemission cut-off (SEPC), as seen in Figure S5E, where $W_f$ = (hν − SEPC) with hν being the energy of the UV photon source (He I line) = 21.22 eV)[9]. Figure S5F presents zoom-in into the UPS spectrum near the Fermi level[10,11]. Here we identify low binding energy (BE) photoemission onset with the valence band maximum (VBM) of the material from which the first electronic transition takes place. Ionization energy (IE) of $FePS_3$ can be calculated using experimental values of the $W_f$ and VBM, according to definition[10,11], IE = $W_f$ + VBM = 4.17 + 0.6 = 4.77 eV.

Remarkably, despite all differences between XPS and UPS techniques (different energy sources, penetration depth, cross-sections for the electronic transitions, etc.), the same values of the VBM (0.6 eV) and the second electronic transition (at 2.2 eV) were obtained, further strengthening the reliability of these measurements.

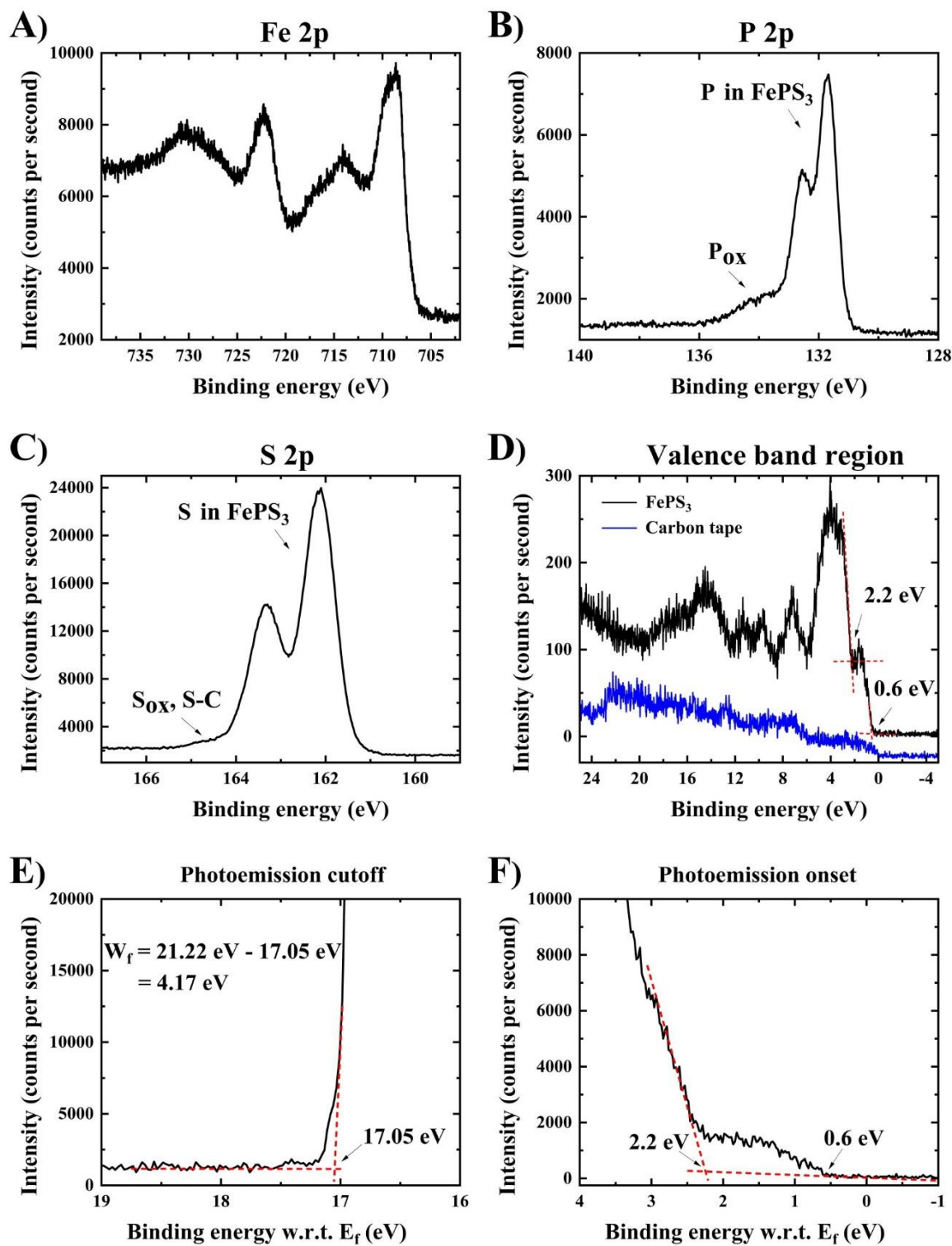

**Figure S5.** XPS and UPS spectroscopies of bulk FePS$_3$: High-resolution XPS spectra of FePS$_3$ showing the (A) Fe 2p, (B) P 2p, (C) S 2p, and (D) Valence band energy regions. XPS spectrum of the carbon tape in the valence band region is shown in (D) (blue line). (E) Close-up of the secondary electron photoemission cut-off (SEPC) region of the FePS$_3$ UPS spectrum, from which the work function value is derived. (F) Zoom-in into the UPS spectrum near the Fermi level (linear intensity scale), revealing the valence band edge (E$_{val}$) energy.

## 5. Experimental Section

<u>FePS$_3$ synthesis</u>

Iron phosphorus trisulfide was synthesized in a quartz tube via chemical vapor transport (CVT)[12–14]. 1 gram of elements mixture: metal iron powder (from Sigma-Aldrich), red phosphorus (from Riedel-de Haën), and elemental sulfur (from Sigma-Aldrich) with Fe : P : S = 1 : 1 : 3.15 molar ratio was grounded in an agate mortar. The 5% addition of sulfur above the 1 : 1 : 3 stoichiometric ratio served as a transport agent. The mixture was placed into quartz ampoule, evacuated to high vacuum (below $3.5 \times 10^{-5}$ Torr) with turbomolecular pump, and finally closed by a flamer. The sealed ampoule was placed inside a previously warmed 3 zone furnace, calibrated that the mixture of elements (the substrate zone) was kept at 850 °C and the deposition zone was 790 °C (see **Figure 1** – top). After 7 days the furnace was turned off and the sample was left to cool down inside. Then the ampoule was opened and only recrystallized FePS$_3$ was collected. It was also weighted for chemical yield calculation against the maximum mass of FePS$_3$ that could be obtained from the used mixture.

<u>Powder X-ray diffraction (PXRD)</u>

PXRD was acquired in parallel beam mode by Rigaku SmartLab 9 kW, copper K-α radiation 1.5406 Å, equipped with Ge 220 monochromator. Two samples were measured. The grounded product of recrystallized FePS$_3$ and a single FePS$_3$ crystal diffractogram was acquired using a PXRD machine by putting the crystal in preferential orientation with the help of a double tilt stage. The results were compared with database: PDF#04-005-1516[1].

<u>Scanning electron microscopy (SEM)</u>

SEM pictures (measured with backscattered electron (BSE) detector) and energy dispersive X-ray spectrum (EDS), measured with Oxford Instruments INCAx-sight SDD detector and processed by INCA (Oxford Instruments) program, were registered by FEI E-SEM Quanta 200 with 20 kV acceleration voltage and the sample was not coated. High-resolution scanning electron microscope (HR-SEM) pictures and EDS maps, for the uncoated sample, were acquired by Zeiss Ultra-Plus FEG-SEM working on 10 kV acceleration voltage. Micrographs were acquired with in-lens and secondary electron detectors. EDS maps presented in Figure 1d, using an Oxford Instruments X-Max 80 mm$^2$ SDD detector, were based on 20 frames and calculated by INCA (Oxford Instruments) program.

<u>Mechanical exfoliation of vdW crystals directly onto a transmission electron microscope (TEM) grid</u>

FePS$_3$ was mechanically exfoliated directly onto a TEM grid by the protocol developed in our laboratory[2]. The described method proved its versatility and has been successfully applied for the exfoliation and transfer of other lamellar compounds like CrPS$_4$, MnPS$_3$, CoPS$_3$, and NiPS$_3$.[2,3]. In the present work, the mechanical exfoliation implemented the use of the tape: "Ultron systems, INC. Silicon-Free Blue Adhesive Film P/N 1009R-6.0", which increased the entire protocol efficiency compared to previous attempts. Copper TEM grids with ultrathin carbon film on lacey carbon (Ted Pella, INC. Ultrathin carbon film on lacey carbon support film, 400 mesh, copper) were used for sample preparation for the TEM measures. Finally, the TEM grid with exfoliated FePS$_3$ was purified via vacuum heating (2 hours in 120 °C) before TEM and STEM investigations took place[2].

<u>Transmission electron microscopy (TEM)</u>

High-resolution (HR) bright-field pictures (BF) and selected area electron diffraction (SAED) patterns were acquired for mechanically exfoliated FePS$_3$ with a transmission electron microscope (TEM): FEI Tecnai G$^2$ T20 S-Twin TEM working at 200 kV acceleration voltage. HR pictures acquired by TEM were filtered with ABSF (average background subtraction filter) with DigitalMicrograph® (Gatan) program.

High-resolution scanning transmission electron microscopy (HR-STEM) with atomically resolved energy dispersive X-ray spectroscopy (EDS) maps

Previously prepared TEM grid of mechanically exfoliated $FePS_3$ was measured in STEM mode with 60 kV acceleration voltage by double-corrected high-resolution scanning/transmission electron microscope (HR-S/TEM) Titan Cubed Themis $G^2$ 60–300 (FEI/Thermo Fisher) equipped with Dual-X detector (Bruker) with an effective solid angle of 1.76 sr for fast and precise local (atomic) chemical analysis. The micrograph presented in **Figure 3a** is based on 100 averaged and aligned frames acquired by the high angle annular dark-field (HAADF) detector. EDS maps (Figure 3b–d) were based on 2000 frames. All EDS maps were pre-filtrated by pixel averaging (3 px) and post-filtrated by radial Wiener filtering (highest frequency 35.0 and edge smoothing 5.0) with Velox Software (Thermo Fisher Scientific). $FePS_3$ was measured with low acceleration voltage, that is 60 kV, to reduce sample damage.

X-ray photoelectron (XPS) and ultraviolet photoemission spectroscopy (UPS)

XPS and UPS measurements were carried out with the Kratos AXIS ULTRA system using a concentric hemispherical analyzer for photo-excited electron detection. UPS was measured with a helium discharge lamp, using He I (21.22 eV) and He II (40.8 eV) radiation lines[9]. The total energy resolution was less than 100 meV, as determined from the Fermi edge of the clean Au reference sample[10,11]. The same Au reference was used for the determination of the Fermi level position ($E_f$) of the instrument (sample), where, by definition[10,11], the binding energy is equal to zero ($E_f = 0$ eV). All UPS spectra were measured with a −10 V bias applied to the sample to observe secondary electron photoemission cut-off at low kinetic energies.

XPS measurements were performed using a monochromatic Al Kα X-ray source (hν = 1486.6 eV) at 75 W and detection pass energies ranging between 20 and 80 eV. Curve fitting analysis was based on linear or Shirley background subtraction and application of Gaussian-Lorentzian line shapes.

Optical investigation

The Raman spectrum was acquired in the range 70–500 cm$^{-1}$ on Horiba Jobin Yvon (LabRAM HR Evolution®) Micro-Raman spectrometer with green laser (532 nm), 1800 gr/mm grating, 50x objective, and at 300 K with the help of a temperature control table.

The photoacoustic spectrum was measured using the gas-microphone method employing an electret condenser microphone mounted in a sealed aluminum cell with a quartz transmission window. The sample was excited with a tunable light source consisting of a 150 W quartz tungsten halogen (QTH) lamp and a 300 mm focal length monochromator. The excitation beam was then mechanically modulated at a frequency of 40 Hz and focused on the sample surface. A detailed description of the photoacoustic experiment can be found in Ref.[15]

Temperature-dependent optical absorption spectra were measured using the same tunable light source as for photoacoustic experiments. The intensity of light transmitted through the sample was measured using Si and InGaAs photodiodes (Hamamatsu) for 1.2–2.2 and 0.6–1.2 eV ranges, respectively. A closed-cycle He cryocooler equipped with a cartridge heater was used for controlling the sample temperature.

Computational details

The calculations are performed in the framework of spin-polarized DFT, using the projector-augmented-wave (PAW)[16] based Vienna ab initio Simulation Package (VASP)[17,18]. The standard exchange-correlation functionals such as LDA, GGA are known to inadequately describe strongly correlated systems which contain transition metals (*3d* states). Therefore, the PBE+U method following the approach of Dudarev et al.[19] for two values of $U_{eff} = U - J$, namely, 2.6 eV and 5.3 eV have been employed, hereafter indicated as U. In the case of the monolayer structure (20 atoms in the cell) the static hybrid functional HSE06 [20] have been performed, on the top of the pre-converged results and optimized structure. A cutoff of 400 eV is chosen for the plane-wave basis set. A k-mesh of $10 \times 6 \times 1$ ($10 \times 6 \times 9$) in the case of monolayer (ML) cell which contains 20 atoms (bulk cell contains 40 atoms) is taken to sample the first Brillouin zone on Γ-centered symmetry reduced Monkhorst-Pack meshes using Gaussian smearing of 0.05 eV. However, in the case of the density of states (DOS), the tetrahedron method was employed along with the denser k-point grids for laterally rectangular supercell $20 \times 12 \times 1$ (ML) and $20 \times 12 \times 9$ (bulk). Both lattice parameters and the position of atoms are relaxed until the maximal force per atom is less than $10^{-3}$ eV/Å, and the maximal component of the stress tensor is less than 0.5 kbar. For the proper description of the monolayer system, the 20 Å of vacuum is added to avoid the spurious interactions between the replicates. However, the bulk $FePS_3$ is a van der Waals layered structure, where the adjacent layers are stacked by the weak van der Waals forces. These forces are crucially important in the proper description of many systems[21–23]. In this work, the vdW forces are taken into account for all calculations employed here, using the method of Grimme[24] with DFT + D3 parametrization[25].

*Structural properties*

The $FePS_3$ is an Ising-like antiferromagnet (AFM) with a honeycomb lattice[26]. The magnetic ground state exhibits the AFM zigzag order within the plane, as well as anti-ferromagnetically aligned adjacent layers[14]. Thus in the case of the monolayer system, the lateral rectangular supercell is used (**Figure S6A**). To include the magnetic ground state for the bulk structure the unit cell is elongated twice in the c direction (**Figure 7a**). The structural parameters are in good agreement with previously reported results[27].

|  | a (Å) | b (Å) | c (Å) | β (°) |
|---|---|---|---|---|
| ML U = 2.6 eV | 5.963 | 10.344 |  |  |
| ML U = 5.3 eV | 5.987 | 10.386 |  |  |
| Bulk U = 2.6 eV | 5.912 | 10.358 | 6.747 | 107.050 |
| Bulk U = 5.3 eV | 6.005 | 10.328 | 6.797 | 107.050 |
| Experiment | 5.947 | 10.300 | 6.722 | 107.160 |

**Table S1**. Lattice parameters obtained for U = 2.6 eV and U = 5.3 eV for monolayer and bulk systems.

*The proper choice of the U parameter*

It is well known that the bandgaps in the standard DFT approach (LDA, GGA exchange-correlation functionals) are underestimated. Therefore, the DFT + U approach is used. However, the DFT + U method is essentially empirical, in the sense that the U parameter must be provided. The proper choice of the U parameter is of fundamental importance to obtain an accurate band structure. Our criterion is to choose U by direct comparison with the hybrid functional HSE06[20] as well as the bandgap obtained in optical measurements; the former is known to give a reasonable prediction of bandgaps of semiconductors. However, as the hybrid functional calculations are computationally very demanding, hence, they are feasible up to several atoms in the supercell. Thus, the comparison is done for the monolayer system, for which the magnetic ground state is an AFM zigzag state[28].

Let us discuss the electronic structure of the monolayer system, by comparing the band projections for U = 2.6 eV, U = 5.3 eV, and the HSE06 functional (Figure S4). The greater the U parameter, the larger bandgap is obtained, however, at the same time the d states are pushed further away from the Fermi level. Thus, it is impossible to choose the parameter U of the 3d states of Fe atoms that will match both the bandgap and the band projections of the hybrid functional in the case of the FePS$_3$ monolayer. On the other hand, the U = 2.6 eV and HSE06 give a similar valence band structure, namely, the small gap between them is visible in the energy range 0.8 eV < E < 0.6 eV. Besides the main contribution of the 3d states of Fe atoms are visible up to 0.6 eV below the Fermi level for both HSE06 and U = 2.6 eV approaches. The HSE06 calculations for the monolayer have been also reported in Ref.[29], however, note that the authors used a primitive cell, which is commensurate for the AFM-Neel magnetic configuration (or FM one), and not adequate for the magnetic ground state of FePS$_3$. The recent report showed that the magnetic arrangement of the spins impacts the electronic structure of MnPS$_3$[28]. In addition, the bands close to the Fermi level are very flat, indicating the atomic-like character of the bands (3d bands of Fe atoms). On the other hand, the U = 5.3 eV gives visible discrepancies in comparison to HSE06 and U = 2.6 eV. The results demonstrate that U = 2.6 eV matches the experimental lattice constant better than U = 5.3 eV (see the results for the bulk in **Figure 8**), and exhibits similar band projections as HSE06 hybrid functional. Thus, the U = 2.6 eV parameter is chosen for further calculations of the bulk FePS$_3$ system. A similar conclusion has been previously obtained in the case of the MnPS$_3$ system[28] and manganese oxides studies[30].

The electronic bandgap obtained in the case of the DFT + D3 + U and the chosen value of U = 2.6 eV matches perfectly the experimental bandgap equal to 1.4 eV of the bulk FePS$_3$. However, the theoretical bandgap of bulk is 0.4 eV smaller than for the ML case (U = 2.6 eV).

VESTA software was used for crystal structure visualization both in the main text and SI[31].

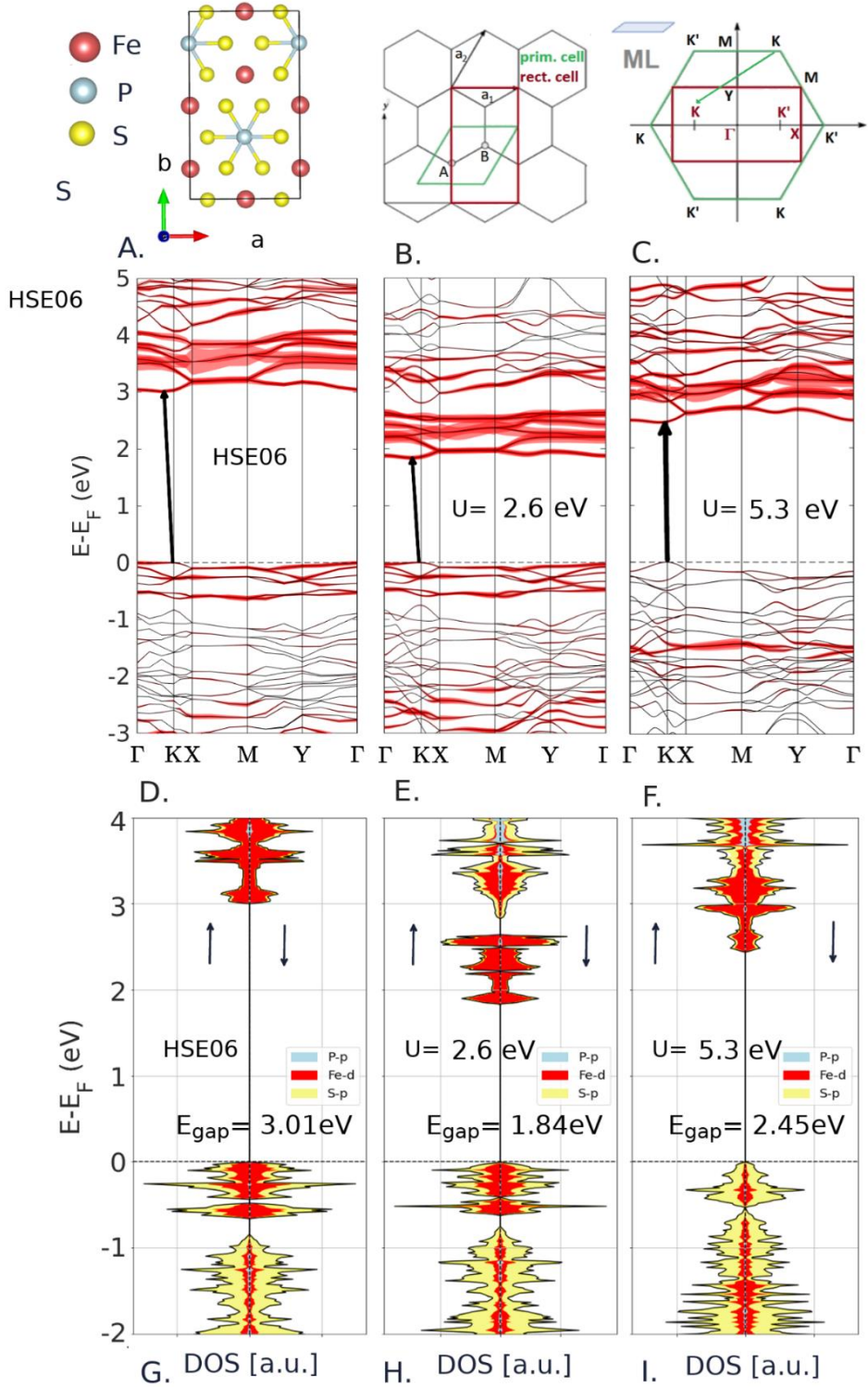

**Figure S6.** Comparison of electronic structure for FePS$_3$ monolayer obtained for U = 2.6 eV, U = 5.3 eV and HSE06. (A) A rectangular lateral supercell is used to capture the magnetic ground state (AFM zigzag). Two possible representations of the hexagonal lattice: (B) primitive unit cell and rectangular unit cell (denoted in red), and (C) their corresponding reciprocal cells in the first Brillouin zone (BZ) are shown. (D–F) The band structure alongside their band projections and (D–F) theirs corresponding projected density of states (PDOS) for HSE06, U = 2.6 eV, U = 5.3 eV approaches are plotted, respectively. The spin-up contribution of the band projections is demonstrated – the spin-down component is the same.

**Supplementary information references**

[1]  G. Ouvrard, R. Brec, J. Rouxel, *Mater. Res. Bull.* **1985**, *20*, 1181.

[2]  A. K. Budniak, N. A. Killilea, S. J. Zelewski, M. Sytnyk, Y. Kauffmann, Y. Amouyal, R. Kudrawiec, W. Heiss, E. Lifshitz, *Small* **2020**, *16*, 1905924.

[3]  M. Shentcis, A. K. Budniak, X. Shi, R. Dahan, Y. Kurman, M. Kalina, H. Herzig Sheinfux, M. Blei, M. K. Svendsen, Y. Amouyal, S. Tongay, K. S. Thygesen, F. H. L. Koppens, E. Lifshitz, F. J. García de Abajo, L. J. Wong, I. Kaminer, *Nat. Photonics* **2020**, *14*, 686.

[4]  A. V. Naumkin, A. Kraust-Vass, S. W. Gaarenstroom, C. J. Powell, NIST X-ray Photoelectron Spectroscopy Database, National Institute of Standards and Technology.

[5]  J. F. Moulder, W. F. Stickle, P. E. Sobol, K. D. Bomben, *Handbook of X-Ray Photoelectron Spectroscopy: A Reference Book of Standard Spectra for Identification and Interpretation of XPS Data*, **1992**.

[6]  Y. Ohno, A. Mineo, I. Matsubara, *Phys. Rev. B* **1989**, *40*, 10262.

[7]  M. Piacentini, F. S. Khumalo, G. Leveque, C. G. Olson, D. W. Lynch, *Chem. Phys.* **1982**, *72*, 61.

[8]  V. Zhukov, S. Alvarez, D. Novikov, *J. Phys. Chem. Solids* **1996**, *57*, 647.

[9]  G. Aeppli, J. J. Donelon, D. E. Eastman, R. W. Johnson, R. A. Pollak, H. J. Stolz, *J. Electron Spectrosc. Relat. Phenom.* **1978**, *14*, 121.

[10] D. Cahen, A. Kahn, *Adv. Mater.* **2003**, *15*, 271.

[11] A. Kahn, *Mater. Horiz.* **2016**, *3*, 7.

[12] W. Klingen, R. Ott, H. Hahn, *ZAAC - J. Inorg. Gen. Chem.* **1973**, *396*, 271.

[13] B. E. Taylor, J. Steger, A. Wold, *J. Solid State Chem.* **1973**, *7*, 461.

[14] R. Brec, *Solid State Ion.* **1986**, *22*, 3.

[15] S. J. Zelewski, R. Kudrawiec, *Sci. Rep.* **2017**, *7*, 1.

[16] D. Joubert, *Phys. Rev. B* **1999**, *59*, 1758.

[17] G. Kresse, J. Hafner, *Phys. Rev. B* **1993**, *48*, 13115.

[18] G. Kresse, J. Furthmüller, *Comput. Mater. Sci.* **1996**, *6*, 15.

[19] S. Dudarev, G. Botton, *Phys. Rev. B* **1998**, *57*, 1505.

[20] A. V. Krukau, O. A. Vydrov, A. F. Izmaylov, G. E. Scuseria, *J. Chem. Phys.* **2006**, *125*, 224106.

[21] M. Birowska, K. Milowska, J. A. Majewski, *Acta Phys. Pol. A* **2011**, *120*, 845.

[22] K. Milowska, M. Birowska, J. A. Majewski, *AIP Conf. Proc.* **2011**, *1399*, 827.

[23] M. Birowska, M. E. Marchwiany, C. Draxl, J. A. Majewski, *Comput. Mater. Sci.* **2021**, *186*, 109940.

[24] S. Grimme, *J. Comput. Chem.* **2006**, *27*, 1787.

[25] S. Grimme, J. Antony, S. Ehrlich, H. Krieg, *J. Chem. Phys.* **2010**, *132*, 154104.

[26] D. Lançon, H. C. Walker, E. Ressouche, B. Ouladdiaf, K. C. Rule, G. J. McIntyre, T. J. Hicks, H. M. Rønnow, A. R. Wildes, *Phys. Rev. B* **2016**, *94*, 214407.

[27] B. L. Chittari, Y. Park, D. Lee, M. Han, A. H. Macdonald, E. Hwang, J. Jung, *Phys. Rev. B* **2016**, *94*, 184428.

[28] M. Birowska, P. E. Faria Junior, J. Fabian, J. Kunstmann, *Phys. Rev. B* **2021**, *103*, L121108.

[29] X. Zhang, X. Zhao, D. Wu, Y. Jing, Z. Zhou, *Adv. Sci.* **2016**, *3*, 1600062.

[30] C. Franchini, R. Podloucky, J. Paier, M. Marsman, G. Kresse, *Phys. Rev. B* **2007**, *75*, 195128.

[31] K. Momma, F. Izumi, *J. Appl. Crystallogr.* **2011**, *44*, 1272.